\documentclass[prl, reprint, preprintnumbers, longbibliography]{revtex4-2}

\usepackage{amsfonts}
\usepackage{amssymb}
\usepackage{amsmath}
\usepackage{graphicx}
\usepackage{dcolumn}
\usepackage{bm}
\usepackage{color}
\usepackage{braket}
\usepackage[breaklinks]{hyperref}
\usepackage{float}
\usepackage{eso-pic}
\usepackage{calc}

\usepackage[utf8]{inputenc}
\begin{document}
	
	\title{Quantum Criticality from Spectral Collapse in the Two-Photon Rabi Model}
	
	\author{Jiong Li$^{1}$}
	\author{Jun-Ling Wang$^{2}$}
	\author{Qing-Hu Chen$^{2,1,}$}
	\email{qhchen@zju.edu.cn}
	\author{Hai-Qing Lin$^{1,}$}
	\email{hqlin@zju.edu.cn}
	
	\affiliation{
		$^1$ Institute for Advanced Study in Physics and School of Physics, Zhejiang University, Hangzhou 310058, China \\
		$^2$ Zhejiang Key Laboratory of Micro-Nano Quantum Chips and Quantum Control, School of Physics, Zhejiang University, Hangzhou 310058, China 
	}
	\date{\today}

	\begin{abstract}
		Spectral collapse in the two-photon quantum Rabi model (tpQRM) has long been regarded as insufficient to establish quantum criticality because, under generic conditions, the lowest excitation gap remains finite. We show that, at a special qubit frequency, spectral collapse in the anisotropic tpQRM constitutes a continuous quantum phase transition governed by a single soft mode. The same-parity gap, set by the vanishing effective oscillator frequency, closes with exponent $z\nu = 1/2$, whereas the lower different-parity gap arises from symmetry-induced splitting. Parity symmetry excludes the different-parity excitation from the quantum Fisher information response and restricts Kibble-Zurek dynamics to excitations within the ground-state parity sector. Together, these results identify the same-parity excitation as the soft mode and its gap as the characteristic energy scale. The corresponding critical exponents, $\nu=1/4$ and $z=2$, place the anisotropic tpQRM in the same universality class as the standard QRM. Our results establish spectral collapse as a potential route to experimentally accessible quantum criticality in a few-body system and show that universality is determined by the soft-mode structure rather than by the closing of the lowest excitation gap.
	\end{abstract}

	\maketitle
	
	\paragraph{Introduction.---}
	The quantum Rabi model (QRM), describing a two-level system coupled to a single quantized mode, provides a paradigmatic framework for light-matter interactions~\cite{rabi_process_1936, scully_quantum_1997, meystre_elements_2007, braak_semi-classical_2016, meystre_quantum_2021}. Advances in circuit quantum electrodynamics (QED)~\cite{wallraff_strong_2004, peropadre_switchable_2010, niemczyk_circuit_2010}, trapped-ion platforms~\cite{pedernales_quantum_2015, lv_quantum_2018, koch_quantum_2023}, and semiconductor quantum systems~\cite{hennessy_quantum_2007, englund_controlling_2007} have enabled experimental access to the ultrastrong and deepstrong coupling regimes, where qualitatively new phenomena emerge beyond the rotating-wave approximation~\cite{wilson_observation_2011, cao_qubit_2011, ridolfo_photon_2012, kena-cohen_ultrastrongly_2013, felicettiUniversalSpectralFeatures2020}. 
	
	A prominent example is the superradiant quantum phase transition (QPT) in the standard QRM, which  emerges in the limit of an infinite qubit-to-mode frequency ratio~\cite{ashhab_superradiance_2013, hwang_quantum_2015}. Although this transition has been experimentally explored and connected to quantum critical phenomena~\cite{hwang_dissipative_2018, cai_observation_2021, chen_experimental_2021, wu_experimental_2024}, it relies on an idealized limit. Realizing quantum criticality in few-body systems under experimentally accessible conditions therefore remains an open challenge.
	
	Two-photon interactions provide a promising route toward this goal~\cite{garbe_superradiant_2017, cui_nonlinear_2020, chen_experimental_2021}. They can be realized in circuit QED~\cite{felicetti_two-photon_2018, felicetti_ultrastrong-coupling_2018, wang_strong_2025} and trapped-ion platforms~\cite{del_valle_two-photon_2010, felicetti_spectral_2015, puebla_quantum_2019}, enabling a wide range of quantum information applications~\cite{casanova_connecting_2018,  villas-boas_multiphoton_2019, piccione_two-photon-interaction_2022}. A hallmark of the two-photon QRM (tpQRM) is \emph{spectral collapse}, where discrete energy levels coalesce into a continuum at a finite coupling strength~\cite{chen_exact_2012, rico_spectral_2020, braak_spectral_2023}. In
	contrast to the QPT in the QRM, spectral collapse instead occurs at finite coupling and finite qubit frequency.
	
	Whether spectral collapse can constitute a continuous QPT, however, remains unresolved. It has long been regarded as a purely spectral phenomenon rather than as a manifestation of quantum criticality~\cite{ying_symmetry-breaking_2021, hammani_two-photon_2024, ying_critical_2025}, because under generic conditions bound states, i.e., normalizable discrete eigenstates below the continuum threshold, remain at the collapse point, leaving the lowest excitation gap finite. Spectral collapse alone does not guarantee the gap closing required for a continuous QPT.
	
	The anisotropic tpQRM provides a notable exception. At a special qubit frequency, the discrete bound-state spectrum terminates at the continuum threshold~\cite{li_critical_2025}, whereas away from this condition, bound states persist and exhibit discrete Efimov-like scaling~\cite{li_critical_2025, zulli_universal_2025}. A complementary approach introduces a quartic stabilizing term that regularizes the spectral-collapse instability and produces a superradiant-like QPT in the limit of an infinite qubit-to-mode frequency ratio~\cite{ying_critical_2025}. These developments leave open the central question addressed here: whether spectral collapse itself, without a stabilizing term or an infinite frequency-ratio limit, can constitute a continuous QPT and, if so, what energy scale governs the resulting universal critical behavior.
	
	In this work, we show that spectral collapse in the anisotropic tpQRM, at the special qubit frequency, indeed constitutes a continuous QPT governed by a single \emph{soft mode}, defined as a low-energy excitation whose gap vanishes continuously at the critical point and sets the characteristic energy scale governing singular critical responses~\cite{belitzHowGenericScale2005, sachdev_quantum_2015, ferriEmergingDissipativePhases2021, stampGallerySoftModes2024}. The low-energy spectrum contains two vanishing excitation gaps with distinct physical origins. The same-parity excitation emerges as the natural soft-mode candidate, with $\epsilon_{\rm sp} \propto |g-g_c|^{z\nu}$ and $z\nu=1/2$, because its gap is set directly by the vanishing effective oscillator frequency. By contrast, the different-parity gap, although the lowest gap in the full spectrum, closes as $\epsilon_{\rm dp}\propto |g-g_c|^{\mu}$ with $\mu=5/4$ and originates from symmetry-induced splitting. Crucially, parity symmetry excludes the different-parity excitation from the quantum Fisher information (QFI) response and restricts the Kibble-Zurek (KZ) dynamics to excitations within the ground-state parity sector. Together, these results identify $\epsilon_{\rm sp}$ as the characteristic energy scale. The quadrature fluctuations then yield the critical exponents $\nu=1/4$ and $z=2$ and characterize the associated critical ground-state response.
	
	These results reveal that universality is determined by the underlying soft-mode structure rather than by the closing of the lowest excitation gap, placing the anisotropic tpQRM in the same universality class as the standard QRM. Spectral collapse therefore provides a mechanism for universal critical behavior and, in principle, an experimentally accessible route to quantum criticality in a few-body system at finite coupling and finite qubit frequency.
	
	\paragraph{Model and symmetry.---} 
	We consider the anisotropic tpQRM described by the Hamiltonian
	\begin{eqnarray}
		H &=& - \frac{\Delta}{2} \sigma_x+\omega a^\dagger a + g \frac{1+r}{2} \sigma_{z} \left( a^2 + a^{\dagger 2} \right) \nonumber \\
		&& + g \frac{1-r}{2} i \sigma_y \left( a^2 - a^{\dagger 2} \right),
	\end{eqnarray}
	where $a$ ($a^\dagger$) is the annihilation (creation) operator of a bosonic mode with frequency $\omega$, $\Delta$ is the qubit frequency, $g$ is the two-photon coupling strength, and $r$ parametrizes the anisotropy. $\sigma_i$ ($i = x, y, z$) are the Pauli matrices. Throughout this work, we set $\hbar = \omega = 1$.
	
	The Hamiltonian possesses a discrete $\mathbb{Z}_4$ symmetry generated by
	\begin{equation}
		\Pi = -\sigma_x \exp\!\left(i \frac{\pi}{2} a^\dagger a\right),
		\quad [\Pi, H] = 0,
		\label{parity}
	\end{equation}
	which decomposes the Hilbert space into four symmetry sectors with eigenvalues $\{\pm 1, \pm i\}$. These sectors can be grouped into even- and odd-photon subspaces labeled by the Bargmann indices $q = 1/4$ and $q = 3/4$, respectively. Within each subspace, a residual $\mathbb{Z}_2$ symmetry distinguishes states by positive or negative parity, providing the natural classification of the low-energy spectrum.
	
	\begin{figure}[tbp]
		\centering
		\includegraphics[width=\linewidth]{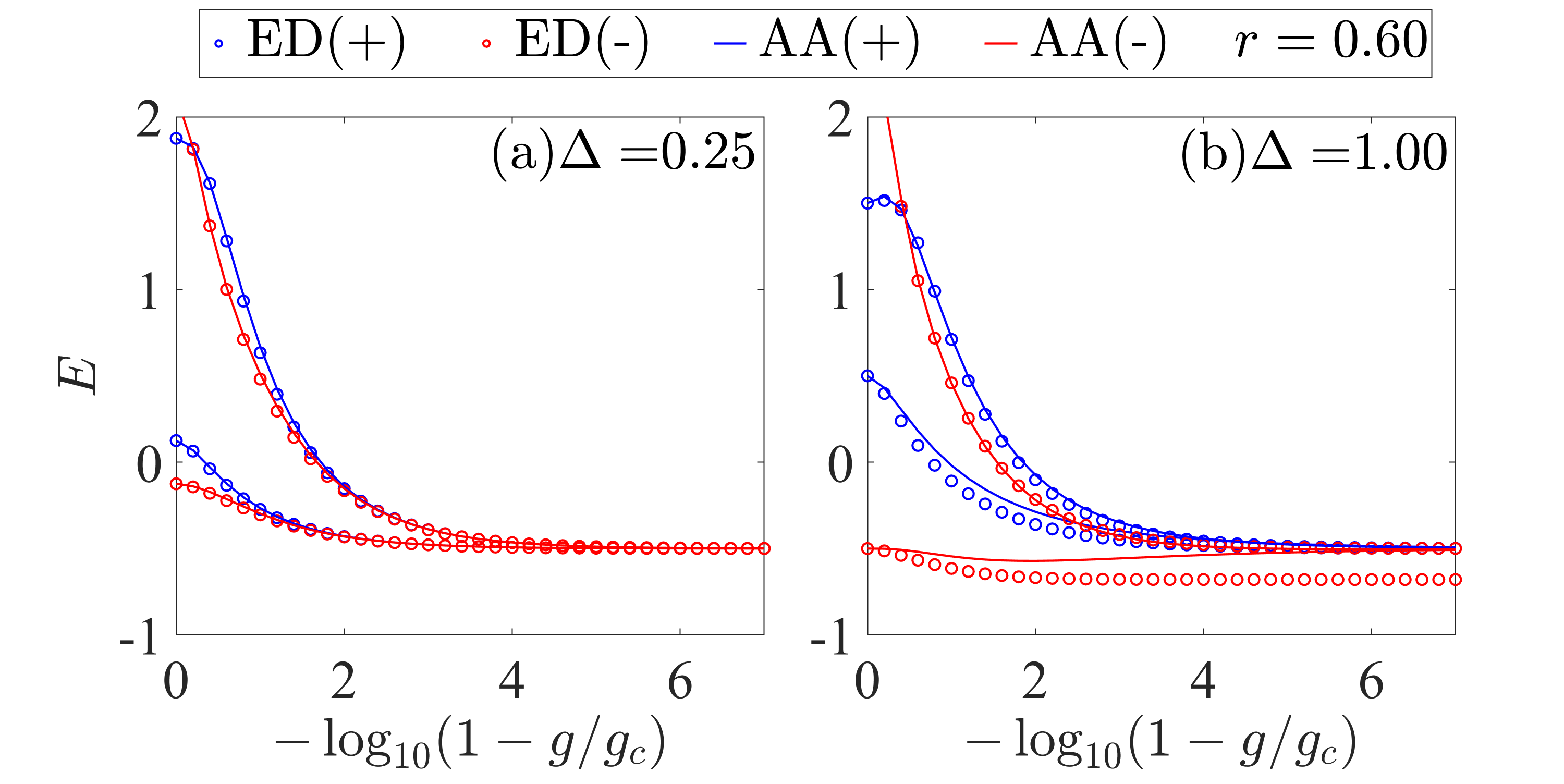}
		\caption{Low-energy spectra in the $q = 1/4$ subspace at $r = 0.60$ for (a) $\Delta = \Delta_c = 0.25$ and (b) $\Delta = 1.0$. The horizontal axis is $-\log_{10}(1 - g/g_c)$. Open circles denote exact diagonalization results, and solid lines denote the adiabatic approximation. Blue and red indicate positive and negative parity, respectively.}
		\label{fig_spectra}
	\end{figure}
	
	As shown in Fig.~\ref{fig_spectra}, the low-energy spectrum exhibits an alternating parity structure near the collapse point. This parity alternation naturally separates the same-parity and different-parity excitation gaps. As we demonstrate below, these two gaps exhibit distinct asymptotic scalings and play fundamentally different roles in the critical behavior.
	
	\paragraph{Low-energy spectral hierarchy.---} 
	We focus on the $q=1/4$ subspace containing the ground state, and first specify the conditions under which spectral collapse gives rise to a continuous QPT. In the anisotropic tpQRM, spectral collapse occurs at
	\begin{equation}
		g_c = \frac{1}{1+r}, 
	\end{equation}
	where the discrete spectrum accumulates at the continuum threshold $E_c=-1/2$~\cite{li_critical_2025}. Spectral collapse alone, however, does not imply a continuous QPT; it does so only under the critical condition
	\begin{equation}
		\Delta = \Delta_c = \frac{1-r}{1+r}, \quad 0<r<1,
	\end{equation}
	under which the discrete bound-state spectrum terminates at the continuum threshold. Away from this condition, discrete bound states remain below the continuum threshold even at $g=g_c$, leaving the excitation gap finite and thereby precluding a continuous QPT.
	
	To analyze the low-energy spectrum near the critical condition, we derive an asymptotically controlled low-energy description using the diagonal adiabatic approximation (AA). A complete derivation is given in the Supplemental Material~\cite{supplement}, which includes Refs.~\cite{xie_quantum_2019, chanBoundStatesTwophoton2020}. Neglecting the couplings between different squeezed-photon manifolds yields the parity eigenstates
	\begin{equation}
		\ket{\psi_{n,\pm}^{(\rm AA)}} = \frac{1}{\sqrt{2}} \begin{bmatrix}
			S(-\theta) \ket{2n} \\
			\mp (-1)^n S(\theta) \ket{2n}
		\end{bmatrix}.
		\label{psi_AA}
	\end{equation}
	Here $S(\theta) = \exp \left[ \frac{\theta}{2} \left( a^{\dagger 2} - a^2 \right)\right]$ is the squeezing operator, with $\theta = \frac{1}{4}\ln [(1+g/g_c)/(1-g/g_c)]$, and the $\mp$ sign corresponds to the parity eigenvalues $\pm 1$. As $g \to g_c$, the AA spectrum admits the asymptotic form 
	\begin{equation}
		E_{n,\pm}^{(\rm AA)} \approx \left(2n + \frac{1}{2} \right) \beta - \frac{1}{2} \pm \frac{K_{nn}}{2} \sqrt{\beta} \left( \delta + \alpha_{nn} \beta^{2} \right),
		\label{E_pm}
	\end{equation}
	where $\delta = \Delta - \Delta_c$ and $\beta = \sqrt{1-g^2/g_c^2} \ll 1$, and $K_{nn}$ and $\alpha_{nn}$ are finite $n$-dependent coefficients. $\beta$ is the effective oscillator frequency: after the squeezing transformation, the two diagonal qubit blocks reduce to the same harmonic-oscillator form with frequency $\beta$. Throughout this work, we approach the collapse point from the confining side $g<g_c$, where $\beta$ remains real. For $g>g_c$, $\beta$ becomes imaginary, indicating the loss of harmonic confinement, and the discrete low-energy spectrum considered here ceases to exist. 
	
	Under the critical condition, Eq.~\eqref{E_pm} becomes asymptotically exact for the low-energy spectrum. As shown in Fig.~\ref{fig_spectra}(a), the AA accurately reproduces the low-energy levels, which approach the continuum threshold $E_c=-1/2$ as $g \to g_c$. For $\delta \neq 0$, however, off-diagonal couplings between different squeezed-photon manifolds become important as $g \to g_c$. Accordingly, Fig.~\ref{fig_spectra}(b) shows clear deviations from the AA: the ground state remains a discrete bound state below the continuum threshold even at $g = g_c$, whereas the low-lying excited states approach the continuum. The leading wavefunction correction scales as $|\psi_{n,\pm}^{(\rm off)}| \propto |\delta| /\sqrt{\beta}$, yielding the crossover scale $\delta^*(\beta) \propto \sqrt{\beta}$. The AA therefore provides a controlled asymptotic low-energy description only when $|\delta| \ll \sqrt{\beta}$~\cite{supplement}.
	
	At $\Delta = \Delta_c$, two low-energy excitation gaps with distinct physical origins emerge. The same-parity gap,
	\begin{equation}
		\epsilon_{\rm sp} = \left| E_{n+1,\pm} - E_{n,\pm} \right| \propto \beta \propto |g - g_c|^{z\nu}, \quad z\nu = 1/2,
	\end{equation}
	originates directly from the vanishing effective oscillator frequency $\beta$, which determines the spacing between adjacent squeezed-photon manifolds. By contrast, the different-parity gap,
	\begin{equation}
		\epsilon_{\rm dp} = |E_{n,+}-E_{n,-}| \propto \beta^{5/2} \propto |g-g_c|^{\mu}, \quad \mu=5/4,
	\end{equation}
	arises from the symmetry-induced splitting between the two parity eigenstates in the same squeezed-photon manifold, rather than directly from the softening of the effective oscillator. Each parity eigenstate consists of two oppositely squeezed oscillator branches. Within a given squeezed-photon manifold, the residual coupling between the two branches produces the symmetry-induced splitting. The corresponding matrix element scales as $\sqrt{\beta} \left( \delta + \alpha_{nn} \beta^2 \right)$, where the overall prefactor $\sqrt{\beta}$ reflects the asymptotically vanishing overlap between the two branches. At $\Delta=\Delta_c$, the term proportional to $\delta$ is absent, so the leading contribution arises from the term $\mathcal{O}(\beta^2)$ in the small-$\beta$ expansion. The splitting therefore scales as $\sqrt{\beta} \times \beta^2 = \beta^{5/2}$, giving $\mu=5/4$. This exponent thus quantifies the asymptotic suppression of the symmetry-induced splitting.
	
	\begin{figure}[tbp]
		\includegraphics[width=1.0\linewidth]{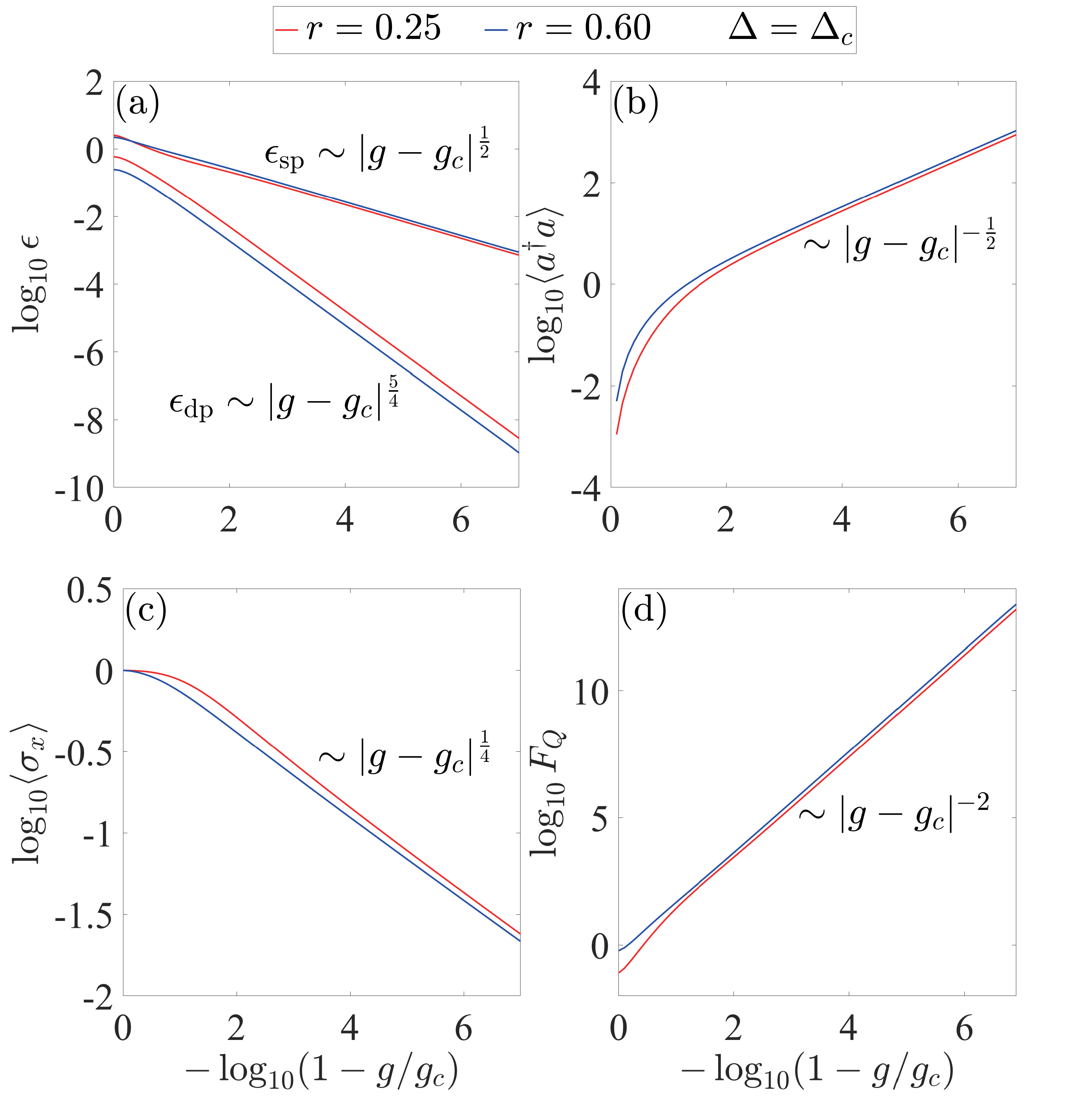}
		\caption{Log-log scaling of (a) the energy gaps $\epsilon_{\rm sp}$ and $\epsilon_{\rm dp}$, (b) the mean photon occupation $\braket{a^\dagger a}$, (c) the atomic polarization $\braket{\sigma_x}$, and (d) the quantum Fisher information $F_Q$ at $\Delta = \Delta_c$ for $r = 0.25$ (red) and $r = 0.60$ (blue). The corresponding scaling exponents are indicated.}
		\label{fig_main}
	\end{figure}

	Beyond the diagonal AA, couplings between different squeezed-photon manifolds generate higher-order energy corrections of order $E_{\rm off} = \mathcal O(\beta^4)$~\cite{supplement}. Collecting the excitation gaps and the leading energy correction gives the asymptotic hierarchy for $g \to g_c$
	\begin{equation}
		E_{\rm off} \propto \beta^4 \quad \ll \quad \epsilon_{\rm dp} \propto \beta^{5/2} \quad \ll \quad \epsilon_{\rm sp} \propto \beta.
	\end{equation}	
	The correction $E_{\rm off}$ is asymptotically subleading to both excitation gaps. Importantly, it is a perturbative correction to the energy levels rather than an excitation gap and therefore does not represent an independent low-energy mode. Exact diagonalization (ED) [Fig.~\ref{fig_main}(a)] confirms the exponents $z\nu = 1/2$ for the same-parity gap and $\mu = 5/4$ for the different-parity gap. 	
	
	In the standard QRM, the corresponding same-parity and different-parity excitation gaps both vanish as $|g-g_c|^{1/2}$~\cite{ashhab_superradiance_2013, hwang_quantum_2015}. In the anisotropic tpQRM, by contrast, they vanish with distinct exponents because they originate from different physical mechanisms. Since $\epsilon_{\rm sp}$ is set directly by the vanishing effective oscillator frequency, while $\epsilon_{\rm dp}$ arises from the symmetry-induced splitting, this distinction motivates the same-parity gap as the natural candidate for the characteristic energy scale, but the spectral hierarchy alone does not establish that the same-parity excitation is the soft mode. This requires determining which excitation controls singular critical responses.
	
	\paragraph{Quantum-information signature.---} 
	The QFI provides a sensitive probe of quantum criticality~\cite{pezze_entanglement_2009,hauke_measuring_2016,garbe_critical_2020,chu_dynamic_2021}. Its spectral representation makes explicit which low-energy excitations contribute to the response:
	\begin{equation}
		F_Q = 4 \sum_{n \neq 0} \frac{\vert \bra{\Phi_n(g)} \partial_g H \ket{\Phi_0 (g)} \vert^2}{\left[ E_n(g) - E_0 (g) \right]^2},
	\end{equation}
	where $H \ket{\Phi_n (g)} = E_n(g) \ket{\Phi_n (g)}$. Since parity symmetry imposes $[\Pi, \partial_g H] = 0$, only excited states with the same parity as the ground state contribute to the QFI. The different-parity excitation is therefore excluded by symmetry, and the dominant contribution arises from the lowest same-parity excitation. 
	
	At $\Delta = \Delta_c$, the QFI diverges as~\cite{supplement}, 
	\begin{equation} 
		F_Q \propto |g-g_c|^{-2} \propto \epsilon_{\rm sp}^{-4},
	\end{equation} 
	derived within the AA and confirmed by ED [Fig.~\ref{fig_main}(d)]. This divergence represents the strongest algebraic enhancement permitted for pure states in Hermitian quantum systems, making the QFI an exceptionally sensitive probe of quantum criticality. 
	
	The symmetry-resolved QFI therefore shows that $\epsilon_{\rm sp}$ is the lowest excitation gap entering this response, whereas $\epsilon_{\rm dp}$ is excluded by parity symmetry.
	
	\paragraph{Kibble-Zurek dynamics.---} 
	Having shown that $\epsilon_{\rm sp}$ is the lowest excitation gap entering the QFI response, we next consider its role in the nonequilibrium dynamics described by the KZ mechanism.
	
	We consider a linear quench $g(t) = g_f t / \tau_q$, ending at $g_f < g_c$ and $g_f / \tau_q \ll 1$. According to KZ theory, adiabatic evolution breaks down near the critical point and generates residual excitations that obey universal scaling laws~\cite{kibble_topology_1976, zurek_cosmological_1985, dziarmaga_dynamics_2010}. Although the applicability of KZ scaling to fully connected systems remains under debate~\cite{acevedo_new_2014, defenu_dynamical_2018}, the anisotropic tpQRM provides a few-body light-matter platform for investigating such dynamics.
	
	\begin{figure}[tbp]
		\includegraphics[width=1.0\linewidth]{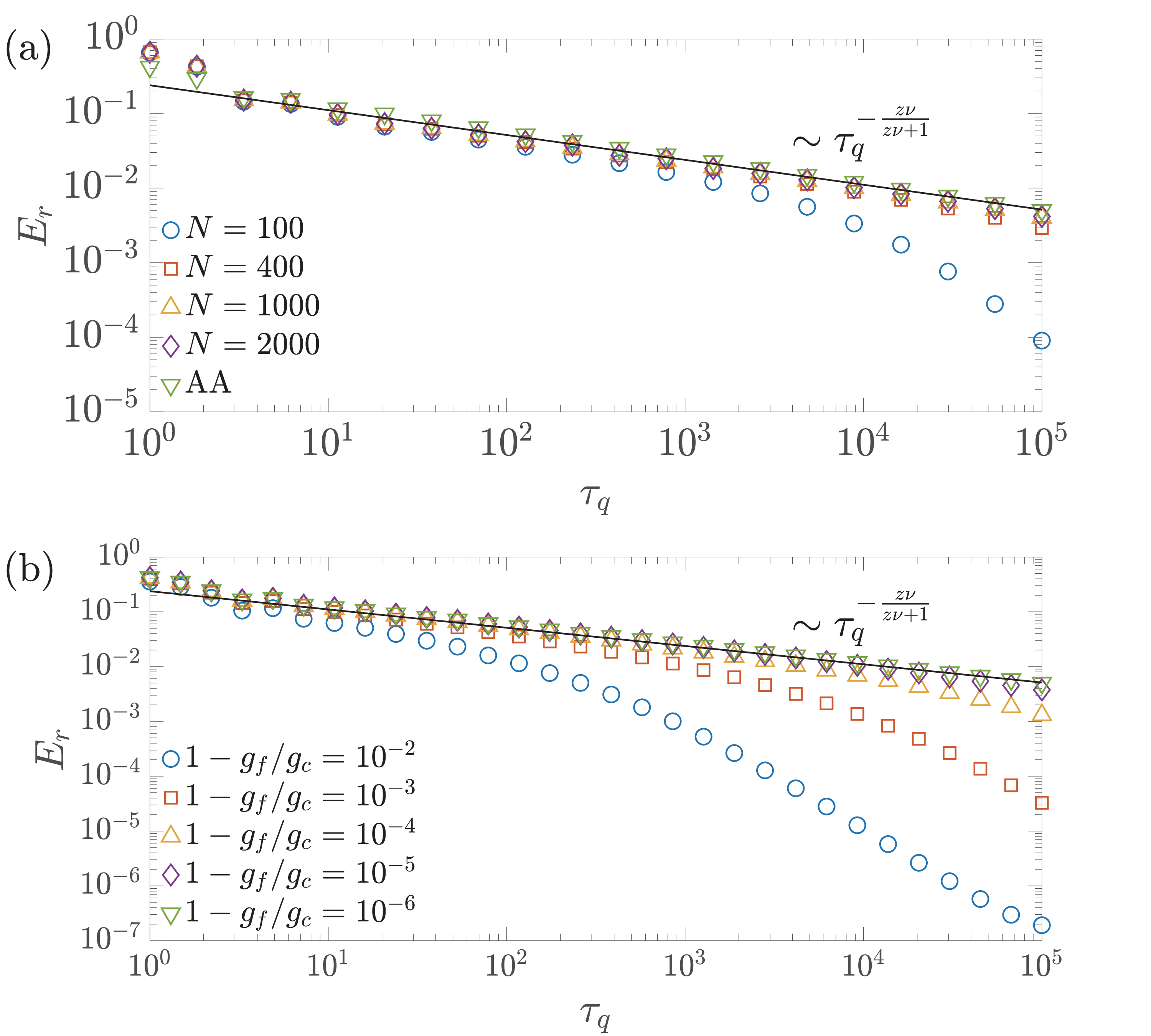}
		\caption{Kibble-Zurek scaling of the residual energy $E_r$. (a) $E_r$ versus the quench time $\tau_q$ for $g_f = (1 - 10^{-6}) g_c$. Exact diagonalization results for different bosonic Fock-space cutoffs $N$ approach the adiabatic approximation result as $N$ increases. (b) Adiabatic approximation results for different final couplings $g_f$. The black line indicates the KZ prediction $E_r \propto \tau_q^{-z\nu/(1+z\nu)} = \tau_q^{-1/3}$.}
		\label{fig_Er}
	\end{figure}
	
	The crossover between the adiabatic and impulsive regimes is determined by the freeze-out condition $\eta^2(g) = \left| \dot{\eta}(g) \right|$. Parity is conserved throughout the quench, so nonadiabatic transitions from the instantaneous ground state are restricted to states with the same parity, and the leading transition is to the lowest same-parity excitation. The  instantaneous gap entering the freeze-out condition is therefore $\eta = \epsilon_{\rm sp}$~\cite{supplement}. With $z\nu = 1/2$, the freeze-out coupling $g_K$ satisfies
	\begin{equation}
		1 - g_K / g_c \propto \tau_q^{-1 / (z\nu + 1)} \propto \tau_q^{-2/3},
	\end{equation}
	which yields the residual energy scaling
	\begin{equation}
		E_r \propto \tau_q^{-\frac{z\nu}{z\nu + 1}} \propto \tau_q^{-1/3}. 
	\end{equation}
	A complete derivation is provided in the Supplemental Material~\cite{supplement}. 
	
	Figure~\ref{fig_Er}(a) demonstrates the numerical convergence trend of the KZ dynamics. As the Fock-space cutoff $N$ increases, the ED results approach the AA prediction, indicating that the observed deviations at large $\tau_q$ originate primarily from finite-cutoff effects. Within the AA, the mean photon occupation satisfies $\braket{a^\dagger a} \approx (2\beta)^{-1}$~\cite{supplement}, providing a practical estimate of the required cutoff. For $1-g_f/g_c=10^{-6}$, this gives $\braket{a^\dagger a} \approx 353$, while the residual-energy curve is already nearly converged at $N=1000$. A cutoff several times larger than the mean photon occupation is therefore sufficient to resolve the KZ regime for the parameters considered here, providing a useful benchmark for the resources required in experimental implementations.
	
	The AA results in Fig.~\ref{fig_Er}(b) further show that the KZ scaling occurs only within an intermediate quench-time window. For fast quenches, the freeze-out energy is no longer well separated from noncritical excitation scales, so higher-energy states contribute appreciably and obscure the KZ scaling. For sufficiently slow quenches ending at $g_f<g_c$, the freeze-out coupling would lie beyond $g_f$, so the protocol terminates while the evolution remains adiabatic, giving the approximate window~\cite{supplement}
	\begin{equation}
		1 \ll \tau_q \ll \left( 1-g_f/g_c \right)^{-3/2},
	\end{equation}
	in units of $\omega=1$ here. As a result, the KZ scaling regime broadens as $g_f\to g_c$, consistent with Fig.~\ref{fig_Er}(b).
	 
	In realistic implementations, dissipation introduces an additional dynamical scale that can truncate the KZ regime. A necessary condition for resolving the KZ scaling is that the relevant dissipation rates remain below the freeze-out soft-mode scale over the experimentally probed regime. Otherwise, dissipation can modify or obscure the closed-system KZ scaling.
	
	The KZ analysis therefore shows that the same-parity excitation also sets the gap relevant to the nonequilibrium dynamics. Together with its origin in the vanishing effective oscillator frequency and its role in the QFI response, this identifies the same-parity excitation as the soft mode, with $\epsilon_{\rm sp}$ setting the characteristic energy scale.
	
	\paragraph{Universality and critical squeezing.---}	
	Having identified the soft mode, we now use the ground-state quadrature scaling to determine the critical exponents. At $\Delta = \Delta_c$, the fluctuations of the quadratures $x=a+a^\dagger$ and $p=i(a^\dagger-a)$ obey~\cite{supplement}
	\begin{equation}
		\Delta x,\Delta p
		\propto \beta^{-1/2}
		\propto |g-g_c|^{-\nu},
		\quad \nu=\frac{1}{4}.
	\end{equation}
	This divergence directly reflects the squeezed-oscillator structure of the low-energy state. Following the finite-component scaling convention used for the standard QRM, this scaling defines $\nu = 1/4$. Together with $z\nu=1/2$, it gives $z=2$. Under the critical condition, these exponents are asymptotically exact and coincide with those of the standard QRM~\cite{ashhab_superradiance_2013, hwang_quantum_2015}, placing the anisotropic tpQRM in the same universality class. The critical exponents remain unchanged throughout the anisotropic regime $0<r<1$. In the isotropic limit $r = 1$ and $\Delta_c = 0$, the different-parity gap vanishes identically, whereas the same-parity gap continues to set the characteristic energy scale~\cite{supplement}. Therefore, the soft-mode structure persists throughout the anisotropic regime $0<r<1$ and extends to the isotropic limit.
	
	\begin{figure}[tbp]
		\includegraphics[width=1.0\linewidth]{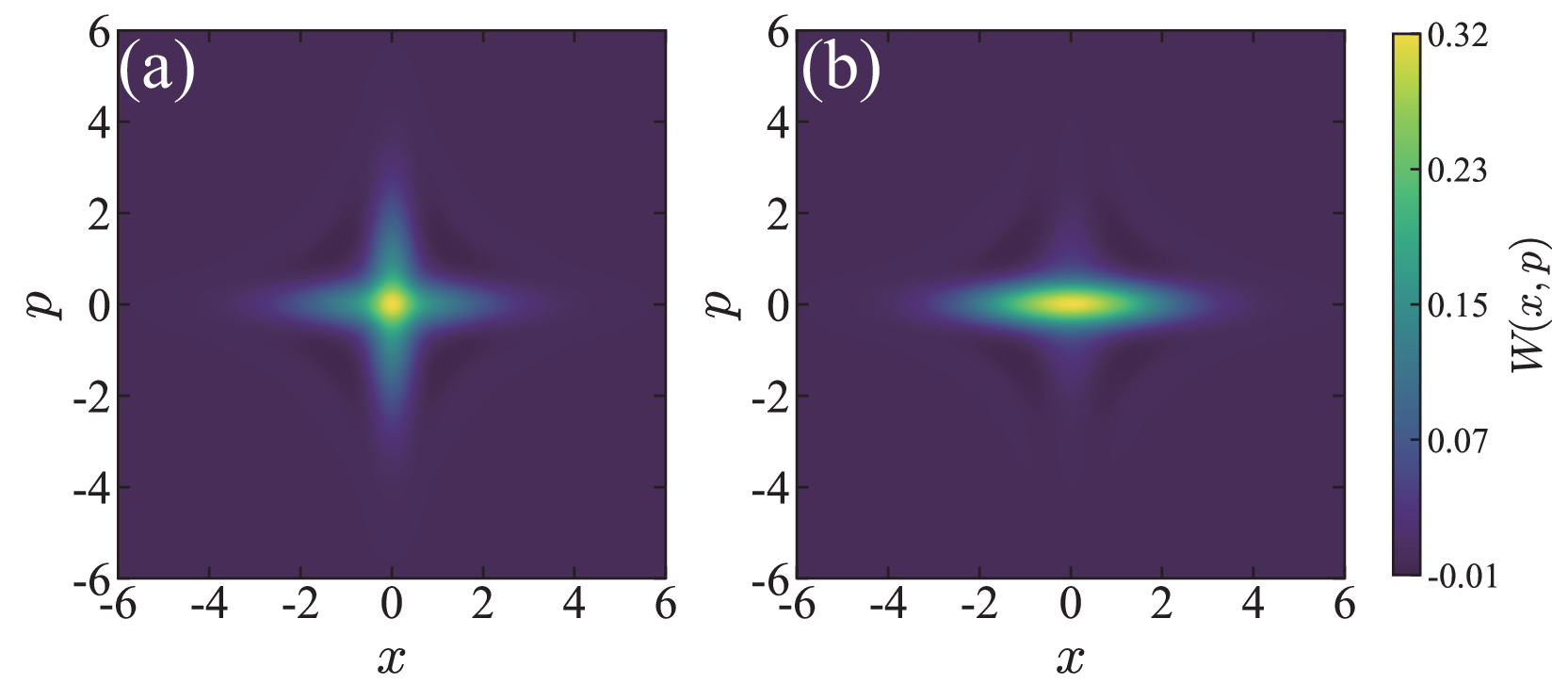}
		\caption{Wigner phase-space distributions $W(x,p)$ of the ground state at $r=0.25$, $\Delta = \Delta_c$, and $g=0.95 g_c$. (a) Reduced $W(x,p)$ obtained by tracing over the qubit. (b) Conditional $W(x,p)$ obtained by projecting onto the $\sigma_z=-1$ eigenstate. The results are computed using QuTiP~\cite{qutip5}.}
		\label{fig_wigner}
	\end{figure}
	
	The associated squeezing structure is visualized in the Wigner phase-space distributions in Fig.~\ref{fig_wigner}, which illustrate the two oppositely squeezed photonic components of the parity eigenstate. Projective measurement of the qubit enables photonic-state postselection: projecting the ground state onto the two $\sigma_z$ eigenstates yields the squeezed vacua $S(-\theta) \ket{0}$ and $S(\theta) \ket{0}$, respectively. This suggests a postselection-based route to generating strongly squeezed states. For each postselected photonic state, $\Delta x \Delta p$ asymptotically approaches the minimum-uncertainty bound, offering a clear experimental signature of the critical squeezing.
	
	The critical ground state also exhibits singular macroscopic responses. At $\Delta=\Delta_c$, the atomic polarization, which serves as the order parameter for the continuous QPT, scales as~\cite{supplement}
	\begin{equation}
		\braket{\sigma_x} 
		\propto |g-g_c|^{1/4}
		\propto \sqrt{\epsilon_{\rm sp}},
	\end{equation}
	and vanishes continuously as the qubit state approaches an equal-weight superposition. By contrast, the mean photon occupation diverges as~\cite{supplement}
	\begin{equation}
		\braket{a^\dagger a} 
		\propto |g-g_c|^{-1/2}
		\propto \epsilon_{\rm sp}^{-1},
	\end{equation}
	reflecting the growth of bosonic fluctuations as the effective oscillator develops increasingly strong squeezing. The ED results for representative anisotropies in Fig.~\ref{fig_main}(b,c) confirm these scaling relations and agree quantitatively with the AA. 
	
	\paragraph{Conclusion.---} 
	We have shown that, under the critical condition, spectral collapse in the anisotropic tpQRM constitutes a continuous QPT governed by a single soft mode. The same-parity gap sets the characteristic energy scale, controlling both equilibrium and nonequilibrium critical behavior, thereby placing the system in the same universality class as the standard QRM~\cite{hwang_quantum_2015, ashhab_superradiance_2013}. By contrast, the different-parity gap originates from the symmetry-induced splitting and, despite being the lowest gap in the full spectrum, does not govern critical responses. These results establish that universality is determined by the underlying soft-mode structure rather than by the closing of the lowest excitation gap.
	
	Beyond this conceptual insight, our work identifies spectral collapse under the critical condition as a mechanism for universal critical behavior at finite coupling and finite qubit frequency, providing, in principle, an experimentally accessible route to quantum criticality in few-body light-matter systems. These results may also enable the preparation of strongly squeezed states and enhanced quantum metrology in nonlinear light-matter platforms operating in the ultrastrong-coupling regime. 
	
	More broadly, our results suggest a natural route toward many-body generalizations. In the single-photon case, continuous QPTs occur in the Dicke model in the thermodynamic limit~\cite{emary_quantum_2003} and in the QRM in an infinite frequency-ratio limit~\cite{hwang_quantum_2015, ashhab_superradiance_2013}. By contrast, although the isotropic two-photon Dicke model exhibits a QPT~\cite{garbe_superradiant_2017, chen_finite-size_2018}, its single-qubit counterpart does not. Our results show that anisotropy enables quantum criticality already at the single-qubit level, suggesting that it may likewise induce continuous QPTs in finite-qubit two-photon Dicke systems without requiring either the thermodynamic limit or an infinite frequency-ratio limit.
	
	\medskip
	\begin{acknowledgments}
		We acknowledge financial support from the National Key R\&D Program of China (Grants No. 2023YFA1406703 and No. 2024YFA1408900) and the National Natural Science Foundation of China under Grant No. 92565201.
	\end{acknowledgments}
	

\begin{thebibliography}{62}%
		\makeatletter
		\providecommand \@ifxundefined [1]{%
			\@ifx{#1\undefined}
		}%
		\providecommand \@ifnum [1]{%
			\ifnum #1\expandafter \@firstoftwo
			\else \expandafter \@secondoftwo
			\fi
		}%
		\providecommand \@ifx [1]{%
			\ifx #1\expandafter \@firstoftwo
			\else \expandafter \@secondoftwo
			\fi
		}%
		\providecommand \natexlab [1]{#1}%
		\providecommand \enquote  [1]{``#1''}%
		\providecommand \bibnamefont  [1]{#1}%
		\providecommand \bibfnamefont [1]{#1}%
		\providecommand \citenamefont [1]{#1}%
		\providecommand \href@noop [0]{\@secondoftwo}%
		\providecommand \href [0]{\begingroup \@sanitize@url \@href}%
		\providecommand \@href[1]{\@@startlink{#1}\@@href}%
		\providecommand \@@href[1]{\endgroup#1\@@endlink}%
		\providecommand \@sanitize@url [0]{\catcode `\\12\catcode `\$12\catcode
			`\&12\catcode `\#12\catcode `\^12\catcode `\_12\catcode `\%12\relax}%
		\providecommand \@@startlink[1]{}%
		\providecommand \@@endlink[0]{}%
		\providecommand \url  [0]{\begingroup\@sanitize@url \@url }%
		\providecommand \@url [1]{\endgroup\@href {#1}{\urlprefix }}%
		\providecommand \urlprefix  [0]{URL }%
		\providecommand \Eprint [0]{\href }%
		\providecommand \doibase [0]{https://doi.org/}%
		\providecommand \selectlanguage [0]{\@gobble}%
		\providecommand \bibinfo  [0]{\@secondoftwo}%
		\providecommand \bibfield  [0]{\@secondoftwo}%
		\providecommand \translation [1]{[#1]}%
		\providecommand \BibitemOpen [0]{}%
		\providecommand \bibitemStop [0]{}%
		\providecommand \bibitemNoStop [0]{.\EOS\space}%
		\providecommand \EOS [0]{\spacefactor3000\relax}%
		\providecommand \BibitemShut  [1]{\csname bibitem#1\endcsname}%
		\let\auto@bib@innerbib\@empty
		\bibitem [{\citenamefont {Rabi}(1936)}]{rabi_process_1936}%
		\BibitemOpen
		\bibfield  {author} {\bibinfo {author} {\bibfnamefont {I.~I.}\ \bibnamefont
				{Rabi}},\ }\bibfield  {title} {\bibinfo {title} {On the process of space
				quantization},\ }\href {https://doi.org/10.1103/PhysRev.49.324} {\bibfield
			{journal} {\bibinfo  {journal} {Phys. Rev.}\ }\textbf {\bibinfo {volume}
				{49}},\ \bibinfo {pages} {324} (\bibinfo {year} {1936})}\BibitemShut
		{NoStop}%
		\bibitem [{\citenamefont {Scully}\ and\ \citenamefont
			{Zubairy}(1997)}]{scully_quantum_1997}%
		\BibitemOpen
		\bibfield  {author} {\bibinfo {author} {\bibfnamefont {M.~O.}\ \bibnamefont
				{Scully}}\ and\ \bibinfo {author} {\bibfnamefont {M.~S.}\ \bibnamefont
				{Zubairy}},\ }\href@noop {} {\emph {\bibinfo {title} {Quantum Optics}}}\
		(\bibinfo  {publisher} {Cambridge University Press},\ \bibinfo {year}
		{1997})\BibitemShut {NoStop}%
		\bibitem [{\citenamefont {Meystre}\ and\ \citenamefont
			{Sargent}(2007)}]{meystre_elements_2007}%
		\BibitemOpen
		\bibfield  {author} {\bibinfo {author} {\bibfnamefont {P.}~\bibnamefont
				{Meystre}}\ and\ \bibinfo {author} {\bibfnamefont {M.}~\bibnamefont
				{Sargent}},\ }\href@noop {} {\emph {\bibinfo {title} {Elements of quantum
					optics}}},\ \bibinfo {edition} {4th}\ ed.\ (\bibinfo  {publisher}
		{Springer},\ \bibinfo {address} {Berlin},\ \bibinfo {year}
		{2007})\BibitemShut {NoStop}%
		\bibitem [{\citenamefont {Braak}\ \emph {et~al.}(2016)\citenamefont {Braak},
			\citenamefont {Chen}, \citenamefont {Batchelor},\ and\ \citenamefont
			{Solano}}]{braak_semi-classical_2016}%
		\BibitemOpen
		\bibfield  {author} {\bibinfo {author} {\bibfnamefont {D.}~\bibnamefont
				{Braak}}, \bibinfo {author} {\bibfnamefont {Q.-H.}\ \bibnamefont {Chen}},
			\bibinfo {author} {\bibfnamefont {M.~T.}\ \bibnamefont {Batchelor}},\ and\
			\bibinfo {author} {\bibfnamefont {E.}~\bibnamefont {Solano}},\ }\bibfield
		{title} {\bibinfo {title} {Semi-classical and quantum {Rabi} models: in
				celebration of 80 years},\ }\href
		{https://doi.org/10.1088/1751-8113/49/30/300301} {\bibfield  {journal}
			{\bibinfo  {journal} {J. Phys. A: Math. Theor.}\ }\textbf {\bibinfo {volume}
				{49}},\ \bibinfo {pages} {300301} (\bibinfo {year} {2016})}\BibitemShut
		{NoStop}%
		\bibitem [{\citenamefont {Meystre}(2021)}]{meystre_quantum_2021}%
		\BibitemOpen
		\bibfield  {author} {\bibinfo {author} {\bibfnamefont {P.}~\bibnamefont
				{Meystre}},\ }\href {https://doi.org/10.1007/978-3-030-76183-7} {\emph
			{\bibinfo {title} {Quantum optics: {Taming} the quantum}}},\ Graduate texts
		in physics\ (\bibinfo  {publisher} {Springer International Publishing},\
		\bibinfo {address} {Cham},\ \bibinfo {year} {2021})\BibitemShut {NoStop}%
		\bibitem [{\citenamefont {Wallraff}\ \emph {et~al.}(2004)\citenamefont
			{Wallraff}, \citenamefont {Schuster}, \citenamefont {Blais}, \citenamefont
			{Frunzio}, \citenamefont {Huang}, \citenamefont {Majer}, \citenamefont
			{Kumar}, \citenamefont {Girvin},\ and\ \citenamefont
			{Schoelkopf}}]{wallraff_strong_2004}%
		\BibitemOpen
		\bibfield  {author} {\bibinfo {author} {\bibfnamefont {A.}~\bibnamefont
				{Wallraff}}, \bibinfo {author} {\bibfnamefont {D.~I.}\ \bibnamefont
				{Schuster}}, \bibinfo {author} {\bibfnamefont {A.}~\bibnamefont {Blais}},
			\bibinfo {author} {\bibfnamefont {L.}~\bibnamefont {Frunzio}}, \bibinfo
			{author} {\bibfnamefont {R.-S.}\ \bibnamefont {Huang}}, \bibinfo {author}
			{\bibfnamefont {J.}~\bibnamefont {Majer}}, \bibinfo {author} {\bibfnamefont
				{S.}~\bibnamefont {Kumar}}, \bibinfo {author} {\bibfnamefont {S.~M.}\
				\bibnamefont {Girvin}},\ and\ \bibinfo {author} {\bibfnamefont {R.~J.}\
				\bibnamefont {Schoelkopf}},\ }\bibfield  {title} {\bibinfo {title} {Strong
				coupling of a single photon to a superconducting qubit using circuit quantum
				electrodynamics},\ }\href {https://doi.org/10.1038/nature02851} {\bibfield
			{journal} {\bibinfo  {journal} {Nature}\ }\textbf {\bibinfo {volume} {431}},\
			\bibinfo {pages} {162} (\bibinfo {year} {2004})}\BibitemShut {NoStop}%
		\bibitem [{\citenamefont {Peropadre}\ \emph {et~al.}(2010)\citenamefont
			{Peropadre}, \citenamefont {Forn-Díaz}, \citenamefont {Solano},\ and\
			\citenamefont {García-Ripoll}}]{peropadre_switchable_2010}%
		\BibitemOpen
		\bibfield  {author} {\bibinfo {author} {\bibfnamefont {B.}~\bibnamefont
				{Peropadre}}, \bibinfo {author} {\bibfnamefont {P.}~\bibnamefont
				{Forn-Díaz}}, \bibinfo {author} {\bibfnamefont {E.}~\bibnamefont {Solano}},\
			and\ \bibinfo {author} {\bibfnamefont {J.~J.}\ \bibnamefont
				{García-Ripoll}},\ }\bibfield  {title} {\bibinfo {title} {Switchable
				ultrastrong coupling in circuit {QED}},\ }\href
		{https://doi.org/10.1103/PhysRevLett.105.023601} {\bibfield  {journal}
			{\bibinfo  {journal} {Phys. Rev. Lett.}\ }\textbf {\bibinfo {volume} {105}},\
			\bibinfo {pages} {023601} (\bibinfo {year} {2010})}\BibitemShut {NoStop}%
		\bibitem [{\citenamefont {Niemczyk}\ \emph {et~al.}(2010)\citenamefont
			{Niemczyk}, \citenamefont {Deppe}, \citenamefont {Huebl}, \citenamefont
			{Menzel}, \citenamefont {Hocke}, \citenamefont {Schwarz}, \citenamefont
			{Garcia-Ripoll}, \citenamefont {Zueco}, \citenamefont {Hümmer},
			\citenamefont {Solano}, \citenamefont {Marx},\ and\ \citenamefont
			{Gross}}]{niemczyk_circuit_2010}%
		\BibitemOpen
		\bibfield  {author} {\bibinfo {author} {\bibfnamefont {T.}~\bibnamefont
				{Niemczyk}}, \bibinfo {author} {\bibfnamefont {F.}~\bibnamefont {Deppe}},
			\bibinfo {author} {\bibfnamefont {H.}~\bibnamefont {Huebl}}, \bibinfo
			{author} {\bibfnamefont {E.~P.}\ \bibnamefont {Menzel}}, \bibinfo {author}
			{\bibfnamefont {F.}~\bibnamefont {Hocke}}, \bibinfo {author} {\bibfnamefont
				{M.~J.}\ \bibnamefont {Schwarz}}, \bibinfo {author} {\bibfnamefont {J.~J.}\
				\bibnamefont {Garcia-Ripoll}}, \bibinfo {author} {\bibfnamefont
				{D.}~\bibnamefont {Zueco}}, \bibinfo {author} {\bibfnamefont
				{T.}~\bibnamefont {Hümmer}}, \bibinfo {author} {\bibfnamefont
				{E.}~\bibnamefont {Solano}}, \bibinfo {author} {\bibfnamefont
				{A.}~\bibnamefont {Marx}},\ and\ \bibinfo {author} {\bibfnamefont
				{R.}~\bibnamefont {Gross}},\ }\bibfield  {title} {\bibinfo {title} {Circuit
				quantum electrodynamics in the ultrastrong-coupling regime},\ }\href
		{https://doi.org/10.1038/nphys1730} {\bibfield  {journal} {\bibinfo
				{journal} {Nat. Phys.}\ }\textbf {\bibinfo {volume} {6}},\ \bibinfo {pages}
			{772} (\bibinfo {year} {2010})}\BibitemShut {NoStop}%
		\bibitem [{\citenamefont {Pedernales}\ \emph {et~al.}(2015)\citenamefont
			{Pedernales}, \citenamefont {Lizuain}, \citenamefont {Felicetti},
			\citenamefont {Romero}, \citenamefont {Lamata},\ and\ \citenamefont
			{Solano}}]{pedernales_quantum_2015}%
		\BibitemOpen
		\bibfield  {author} {\bibinfo {author} {\bibfnamefont {J.~S.}\ \bibnamefont
				{Pedernales}}, \bibinfo {author} {\bibfnamefont {I.}~\bibnamefont {Lizuain}},
			\bibinfo {author} {\bibfnamefont {S.}~\bibnamefont {Felicetti}}, \bibinfo
			{author} {\bibfnamefont {G.}~\bibnamefont {Romero}}, \bibinfo {author}
			{\bibfnamefont {L.}~\bibnamefont {Lamata}},\ and\ \bibinfo {author}
			{\bibfnamefont {E.}~\bibnamefont {Solano}},\ }\bibfield  {title} {\bibinfo
			{title} {Quantum {Rabi} model with trapped ions},\ }\href
		{https://doi.org/10.1038/srep15472} {\bibfield  {journal} {\bibinfo
				{journal} {Sci. Rep.}\ }\textbf {\bibinfo {volume} {5}},\ \bibinfo {pages}
			{15472} (\bibinfo {year} {2015})}\BibitemShut {NoStop}%
		\bibitem [{\citenamefont {Lv}\ \emph {et~al.}(2018)\citenamefont {Lv},
			\citenamefont {An}, \citenamefont {Liu}, \citenamefont {Zhang}, \citenamefont
			{Pedernales}, \citenamefont {Lamata}, \citenamefont {Solano},\ and\
			\citenamefont {Kim}}]{lv_quantum_2018}%
		\BibitemOpen
		\bibfield  {author} {\bibinfo {author} {\bibfnamefont {D.}~\bibnamefont
				{Lv}}, \bibinfo {author} {\bibfnamefont {S.}~\bibnamefont {An}}, \bibinfo
			{author} {\bibfnamefont {Z.}~\bibnamefont {Liu}}, \bibinfo {author}
			{\bibfnamefont {J.-N.}\ \bibnamefont {Zhang}}, \bibinfo {author}
			{\bibfnamefont {J.~S.}\ \bibnamefont {Pedernales}}, \bibinfo {author}
			{\bibfnamefont {L.}~\bibnamefont {Lamata}}, \bibinfo {author} {\bibfnamefont
				{E.}~\bibnamefont {Solano}},\ and\ \bibinfo {author} {\bibfnamefont
				{K.}~\bibnamefont {Kim}},\ }\bibfield  {title} {\bibinfo {title} {Quantum
				simulation of the quantum {Rabi} model in a trapped ion},\ }\href
		{https://doi.org/10.1103/PhysRevX.8.021027} {\bibfield  {journal} {\bibinfo
				{journal} {Phys. Rev. X}\ }\textbf {\bibinfo {volume} {8}},\ \bibinfo {pages}
			{021027} (\bibinfo {year} {2018})}\BibitemShut {NoStop}%
		\bibitem [{\citenamefont {Koch}\ \emph {et~al.}(2023)\citenamefont {Koch},
			\citenamefont {Hunanyan}, \citenamefont {Ockenfels}, \citenamefont {Rico},
			\citenamefont {Solano},\ and\ \citenamefont {Weitz}}]{koch_quantum_2023}%
		\BibitemOpen
		\bibfield  {author} {\bibinfo {author} {\bibfnamefont {J.}~\bibnamefont
				{Koch}}, \bibinfo {author} {\bibfnamefont {G.~R.}\ \bibnamefont {Hunanyan}},
			\bibinfo {author} {\bibfnamefont {T.}~\bibnamefont {Ockenfels}}, \bibinfo
			{author} {\bibfnamefont {E.}~\bibnamefont {Rico}}, \bibinfo {author}
			{\bibfnamefont {E.}~\bibnamefont {Solano}},\ and\ \bibinfo {author}
			{\bibfnamefont {M.}~\bibnamefont {Weitz}},\ }\bibfield  {title} {\bibinfo
			{title} {Quantum {Rabi} dynamics of trapped atoms far in the deep strong
				coupling regime},\ }\href {https://doi.org/10.1038/s41467-023-36611-z}
		{\bibfield  {journal} {\bibinfo  {journal} {Nat. Commun.}\ }\textbf {\bibinfo
				{volume} {14}},\ \bibinfo {pages} {954} (\bibinfo {year} {2023})}\BibitemShut
		{NoStop}%
		\bibitem [{\citenamefont {Hennessy}\ \emph {et~al.}(2007)\citenamefont
			{Hennessy}, \citenamefont {Badolato}, \citenamefont {Winger}, \citenamefont
			{Gerace}, \citenamefont {Atatüre}, \citenamefont {Gulde}, \citenamefont
			{Fält}, \citenamefont {Hu},\ and\ \citenamefont
			{Imamoğlu}}]{hennessy_quantum_2007}%
		\BibitemOpen
		\bibfield  {author} {\bibinfo {author} {\bibfnamefont {K.}~\bibnamefont
				{Hennessy}}, \bibinfo {author} {\bibfnamefont {A.}~\bibnamefont {Badolato}},
			\bibinfo {author} {\bibfnamefont {M.}~\bibnamefont {Winger}}, \bibinfo
			{author} {\bibfnamefont {D.}~\bibnamefont {Gerace}}, \bibinfo {author}
			{\bibfnamefont {M.}~\bibnamefont {Atatüre}}, \bibinfo {author}
			{\bibfnamefont {S.}~\bibnamefont {Gulde}}, \bibinfo {author} {\bibfnamefont
				{S.}~\bibnamefont {Fält}}, \bibinfo {author} {\bibfnamefont {E.~L.}\
				\bibnamefont {Hu}},\ and\ \bibinfo {author} {\bibfnamefont {A.}~\bibnamefont
				{Imamoğlu}},\ }\bibfield  {title} {\bibinfo {title} {Quantum nature of a
				strongly coupled single quantum dot–cavity system},\ }\href
		{https://doi.org/10.1038/nature05586} {\bibfield  {journal} {\bibinfo
				{journal} {Nature}\ }\textbf {\bibinfo {volume} {445}},\ \bibinfo {pages}
			{896} (\bibinfo {year} {2007})}\BibitemShut {NoStop}%
		\bibitem [{\citenamefont {Englund}\ \emph {et~al.}(2007)\citenamefont
			{Englund}, \citenamefont {Faraon}, \citenamefont {Fushman}, \citenamefont
			{Stoltz}, \citenamefont {Petroff},\ and\ \citenamefont
			{Vučković}}]{englund_controlling_2007}%
		\BibitemOpen
		\bibfield  {author} {\bibinfo {author} {\bibfnamefont {D.}~\bibnamefont
				{Englund}}, \bibinfo {author} {\bibfnamefont {A.}~\bibnamefont {Faraon}},
			\bibinfo {author} {\bibfnamefont {I.}~\bibnamefont {Fushman}}, \bibinfo
			{author} {\bibfnamefont {N.}~\bibnamefont {Stoltz}}, \bibinfo {author}
			{\bibfnamefont {P.}~\bibnamefont {Petroff}},\ and\ \bibinfo {author}
			{\bibfnamefont {J.}~\bibnamefont {Vučković}},\ }\bibfield  {title}
		{\bibinfo {title} {Controlling cavity reflectivity with a single quantum
				dot},\ }\href {https://doi.org/10.1038/nature06234} {\bibfield  {journal}
			{\bibinfo  {journal} {Nature}\ }\textbf {\bibinfo {volume} {450}},\ \bibinfo
			{pages} {857} (\bibinfo {year} {2007})}\BibitemShut {NoStop}%
		\bibitem [{\citenamefont {Wilson}\ \emph {et~al.}(2011)\citenamefont {Wilson},
			\citenamefont {Johansson}, \citenamefont {Pourkabirian}, \citenamefont
			{Simoen}, \citenamefont {Johansson}, \citenamefont {Duty}, \citenamefont
			{Nori},\ and\ \citenamefont {Delsing}}]{wilson_observation_2011}%
		\BibitemOpen
		\bibfield  {author} {\bibinfo {author} {\bibfnamefont {C.~M.}\ \bibnamefont
				{Wilson}}, \bibinfo {author} {\bibfnamefont {G.}~\bibnamefont {Johansson}},
			\bibinfo {author} {\bibfnamefont {A.}~\bibnamefont {Pourkabirian}}, \bibinfo
			{author} {\bibfnamefont {M.}~\bibnamefont {Simoen}}, \bibinfo {author}
			{\bibfnamefont {J.~R.}\ \bibnamefont {Johansson}}, \bibinfo {author}
			{\bibfnamefont {T.}~\bibnamefont {Duty}}, \bibinfo {author} {\bibfnamefont
				{F.}~\bibnamefont {Nori}},\ and\ \bibinfo {author} {\bibfnamefont
				{P.}~\bibnamefont {Delsing}},\ }\bibfield  {title} {\bibinfo {title}
			{Observation of the dynamical {Casimir} effect in a superconducting
				circuit},\ }\href {https://doi.org/10.1038/nature10561} {\bibfield  {journal}
			{\bibinfo  {journal} {Nature}\ }\textbf {\bibinfo {volume} {479}},\ \bibinfo
			{pages} {376} (\bibinfo {year} {2011})}\BibitemShut {NoStop}%
		\bibitem [{\citenamefont {Cao}\ \emph {et~al.}(2011)\citenamefont {Cao},
			\citenamefont {You}, \citenamefont {Zheng},\ and\ \citenamefont
			{Nori}}]{cao_qubit_2011}%
		\BibitemOpen
		\bibfield  {author} {\bibinfo {author} {\bibfnamefont {X.}~\bibnamefont
				{Cao}}, \bibinfo {author} {\bibfnamefont {J.~Q.}\ \bibnamefont {You}},
			\bibinfo {author} {\bibfnamefont {H.}~\bibnamefont {Zheng}},\ and\ \bibinfo
			{author} {\bibfnamefont {F.}~\bibnamefont {Nori}},\ }\bibfield  {title}
		{\bibinfo {title} {A qubit strongly coupled to a resonant cavity: asymmetry
				of the spontaneous emission spectrum beyond the rotating wave
				approximation},\ }\href {https://doi.org/10.1088/1367-2630/13/7/073002}
		{\bibfield  {journal} {\bibinfo  {journal} {New J. Phys.}\ }\textbf {\bibinfo
				{volume} {13}},\ \bibinfo {pages} {073002} (\bibinfo {year}
			{2011})}\BibitemShut {NoStop}%
		\bibitem [{\citenamefont {Ridolfo}\ \emph {et~al.}(2012)\citenamefont
			{Ridolfo}, \citenamefont {Leib}, \citenamefont {Savasta},\ and\ \citenamefont
			{Hartmann}}]{ridolfo_photon_2012}%
		\BibitemOpen
		\bibfield  {author} {\bibinfo {author} {\bibfnamefont {A.}~\bibnamefont
				{Ridolfo}}, \bibinfo {author} {\bibfnamefont {M.}~\bibnamefont {Leib}},
			\bibinfo {author} {\bibfnamefont {S.}~\bibnamefont {Savasta}},\ and\ \bibinfo
			{author} {\bibfnamefont {M.~J.}\ \bibnamefont {Hartmann}},\ }\bibfield
		{title} {\bibinfo {title} {Photon blockade in the ultrastrong coupling
				regime},\ }\href {https://doi.org/10.1103/PhysRevLett.109.193602} {\bibfield
			{journal} {\bibinfo  {journal} {Phys. Rev. Lett.}\ }\textbf {\bibinfo
				{volume} {109}},\ \bibinfo {pages} {193602} (\bibinfo {year}
			{2012})}\BibitemShut {NoStop}%
		\bibitem [{\citenamefont {Kéna-Cohen}\ \emph {et~al.}(2013)\citenamefont
			{Kéna-Cohen}, \citenamefont {Maier},\ and\ \citenamefont
			{Bradley}}]{kena-cohen_ultrastrongly_2013}%
		\BibitemOpen
		\bibfield  {author} {\bibinfo {author} {\bibfnamefont {S.}~\bibnamefont
				{Kéna-Cohen}}, \bibinfo {author} {\bibfnamefont {S.~A.}\ \bibnamefont
				{Maier}},\ and\ \bibinfo {author} {\bibfnamefont {D.~D.~C.}\ \bibnamefont
				{Bradley}},\ }\bibfield  {title} {\bibinfo {title} {Ultrastrongly coupled
				exciton–polaritons in metal-clad organic semiconductor microcavities},\
		}\href {https://doi.org/10.1002/adom.201300256} {\bibfield  {journal}
			{\bibinfo  {journal} {Adv. Opt. Mater.}\ }\textbf {\bibinfo {volume} {1}},\
			\bibinfo {pages} {827} (\bibinfo {year} {2013})}\BibitemShut {NoStop}%
		\bibitem [{\citenamefont {Felicetti}\ and\ \citenamefont
			{Le~Boité}(2020)}]{felicettiUniversalSpectralFeatures2020}%
		\BibitemOpen
		\bibfield  {author} {\bibinfo {author} {\bibfnamefont {S.}~\bibnamefont
				{Felicetti}}\ and\ \bibinfo {author} {\bibfnamefont {A.}~\bibnamefont
				{Le~Boité}},\ }\bibfield  {title} {\bibinfo {title} {Universal spectral
				features of ultrastrongly coupled systems},\ }\href
		{https://doi.org/10.1103/PhysRevLett.124.040404} {\bibfield  {journal}
			{\bibinfo  {journal} {Phys. Rev. Lett.}\ }\textbf {\bibinfo {volume} {124}},\
			\bibinfo {pages} {040404} (\bibinfo {year} {2020})}\BibitemShut {NoStop}%
		\bibitem [{\citenamefont {Ashhab}(2013)}]{ashhab_superradiance_2013}%
		\BibitemOpen
		\bibfield  {author} {\bibinfo {author} {\bibfnamefont {S.}~\bibnamefont
				{Ashhab}},\ }\bibfield  {title} {\bibinfo {title} {Superradiance transition
				in a system with a single qubit and a single oscillator},\ }\href
		{https://doi.org/10.1103/PhysRevA.87.013826} {\bibfield  {journal} {\bibinfo
				{journal} {Phys. Rev. A}\ }\textbf {\bibinfo {volume} {87}},\ \bibinfo
			{pages} {013826} (\bibinfo {year} {2013})}\BibitemShut {NoStop}%
		\bibitem [{\citenamefont {Hwang}\ \emph {et~al.}(2015)\citenamefont {Hwang},
			\citenamefont {Puebla},\ and\ \citenamefont {Plenio}}]{hwang_quantum_2015}%
		\BibitemOpen
		\bibfield  {author} {\bibinfo {author} {\bibfnamefont {M.-J.}\ \bibnamefont
				{Hwang}}, \bibinfo {author} {\bibfnamefont {R.}~\bibnamefont {Puebla}},\ and\
			\bibinfo {author} {\bibfnamefont {M.~B.}\ \bibnamefont {Plenio}},\ }\bibfield
		{title} {\bibinfo {title} {Quantum phase transition and universal dynamics
				in the {Rabi} model},\ }\href
		{https://doi.org/10.1103/PhysRevLett.115.180404} {\bibfield  {journal}
			{\bibinfo  {journal} {Phys. Rev. Lett.}\ }\textbf {\bibinfo {volume} {115}},\
			\bibinfo {pages} {180404} (\bibinfo {year} {2015})}\BibitemShut {NoStop}%
		\bibitem [{\citenamefont {Hwang}\ \emph {et~al.}(2018)\citenamefont {Hwang},
			\citenamefont {Rabl},\ and\ \citenamefont {Plenio}}]{hwang_dissipative_2018}%
		\BibitemOpen
		\bibfield  {author} {\bibinfo {author} {\bibfnamefont {M.-J.}\ \bibnamefont
				{Hwang}}, \bibinfo {author} {\bibfnamefont {P.}~\bibnamefont {Rabl}},\ and\
			\bibinfo {author} {\bibfnamefont {M.~B.}\ \bibnamefont {Plenio}},\ }\bibfield
		{title} {\bibinfo {title} {Dissipative phase transition in the open quantum
				{Rabi} model},\ }\href {https://doi.org/10.1103/PhysRevA.97.013825}
		{\bibfield  {journal} {\bibinfo  {journal} {Phys. Rev. A}\ }\textbf {\bibinfo
				{volume} {97}},\ \bibinfo {pages} {013825} (\bibinfo {year}
			{2018})}\BibitemShut {NoStop}%
		\bibitem [{\citenamefont {Cai}\ \emph {et~al.}(2021)\citenamefont {Cai},
			\citenamefont {Liu}, \citenamefont {Zhao}, \citenamefont {Wu}, \citenamefont
			{Mei}, \citenamefont {Jiang}, \citenamefont {He}, \citenamefont {Zhang},
			\citenamefont {Zhou},\ and\ \citenamefont {Duan}}]{cai_observation_2021}%
		\BibitemOpen
		\bibfield  {author} {\bibinfo {author} {\bibfnamefont {M.-L.}\ \bibnamefont
				{Cai}}, \bibinfo {author} {\bibfnamefont {Z.-D.}\ \bibnamefont {Liu}},
			\bibinfo {author} {\bibfnamefont {W.-D.}\ \bibnamefont {Zhao}}, \bibinfo
			{author} {\bibfnamefont {Y.-K.}\ \bibnamefont {Wu}}, \bibinfo {author}
			{\bibfnamefont {Q.-X.}\ \bibnamefont {Mei}}, \bibinfo {author} {\bibfnamefont
				{Y.}~\bibnamefont {Jiang}}, \bibinfo {author} {\bibfnamefont
				{L.}~\bibnamefont {He}}, \bibinfo {author} {\bibfnamefont {X.}~\bibnamefont
				{Zhang}}, \bibinfo {author} {\bibfnamefont {Z.-C.}\ \bibnamefont {Zhou}},\
			and\ \bibinfo {author} {\bibfnamefont {L.-M.}\ \bibnamefont {Duan}},\
		}\bibfield  {title} {\bibinfo {title} {Observation of a quantum phase
				transition in the quantum {Rabi} model with a single trapped ion},\ }\href
		{https://doi.org/10.1038/s41467-021-21425-8} {\bibfield  {journal} {\bibinfo
				{journal} {Nat. Commun.}\ }\textbf {\bibinfo {volume} {12}},\ \bibinfo
			{pages} {1126} (\bibinfo {year} {2021})}\BibitemShut {NoStop}%
		\bibitem [{\citenamefont {Chen}\ \emph {et~al.}(2021)\citenamefont {Chen},
			\citenamefont {Wu}, \citenamefont {Jiang}, \citenamefont {Lü}, \citenamefont
			{Peng},\ and\ \citenamefont {Du}}]{chen_experimental_2021}%
		\BibitemOpen
		\bibfield  {author} {\bibinfo {author} {\bibfnamefont {X.}~\bibnamefont
				{Chen}}, \bibinfo {author} {\bibfnamefont {Z.}~\bibnamefont {Wu}}, \bibinfo
			{author} {\bibfnamefont {M.}~\bibnamefont {Jiang}}, \bibinfo {author}
			{\bibfnamefont {X.-Y.}\ \bibnamefont {Lü}}, \bibinfo {author} {\bibfnamefont
				{X.}~\bibnamefont {Peng}},\ and\ \bibinfo {author} {\bibfnamefont
				{J.}~\bibnamefont {Du}},\ }\bibfield  {title} {\bibinfo {title} {Experimental
				quantum simulation of superradiant phase transition beyond no-go theorem via
				antisqueezing},\ }\href {https://doi.org/10.1038/s41467-021-26573-5}
		{\bibfield  {journal} {\bibinfo  {journal} {Nat. Commun.}\ }\textbf {\bibinfo
				{volume} {12}},\ \bibinfo {pages} {6281} (\bibinfo {year}
			{2021})}\BibitemShut {NoStop}%
		\bibitem [{\citenamefont {Wu}\ \emph {et~al.}(2024)\citenamefont {Wu},
			\citenamefont {Hu}, \citenamefont {Wang}, \citenamefont {Chen}, \citenamefont
			{Li}, \citenamefont {Zhao}, \citenamefont {Lü},\ and\ \citenamefont
			{Peng}}]{wu_experimental_2024}%
		\BibitemOpen
		\bibfield  {author} {\bibinfo {author} {\bibfnamefont {Z.}~\bibnamefont
				{Wu}}, \bibinfo {author} {\bibfnamefont {C.}~\bibnamefont {Hu}}, \bibinfo
			{author} {\bibfnamefont {T.}~\bibnamefont {Wang}}, \bibinfo {author}
			{\bibfnamefont {Y.}~\bibnamefont {Chen}}, \bibinfo {author} {\bibfnamefont
				{Y.}~\bibnamefont {Li}}, \bibinfo {author} {\bibfnamefont {L.}~\bibnamefont
				{Zhao}}, \bibinfo {author} {\bibfnamefont {X.-Y.}\ \bibnamefont {Lü}},\ and\
			\bibinfo {author} {\bibfnamefont {X.}~\bibnamefont {Peng}},\ }\bibfield
		{title} {\bibinfo {title} {Experimental quantum simulation of
				multicriticality in closed and open {Rabi} model},\ }\href
		{https://doi.org/10.1103/PhysRevLett.133.173602} {\bibfield  {journal}
			{\bibinfo  {journal} {Phys. Rev. Lett.}\ }\textbf {\bibinfo {volume} {133}},\
			\bibinfo {pages} {173602} (\bibinfo {year} {2024})}\BibitemShut {NoStop}%
		\bibitem [{\citenamefont {Garbe}\ \emph {et~al.}(2017)\citenamefont {Garbe},
			\citenamefont {Egusquiza}, \citenamefont {Solano}, \citenamefont {Ciuti},
			\citenamefont {Coudreau}, \citenamefont {Milman},\ and\ \citenamefont
			{Felicetti}}]{garbe_superradiant_2017}%
		\BibitemOpen
		\bibfield  {author} {\bibinfo {author} {\bibfnamefont {L.}~\bibnamefont
				{Garbe}}, \bibinfo {author} {\bibfnamefont {I.~L.}\ \bibnamefont
				{Egusquiza}}, \bibinfo {author} {\bibfnamefont {E.}~\bibnamefont {Solano}},
			\bibinfo {author} {\bibfnamefont {C.}~\bibnamefont {Ciuti}}, \bibinfo
			{author} {\bibfnamefont {T.}~\bibnamefont {Coudreau}}, \bibinfo {author}
			{\bibfnamefont {P.}~\bibnamefont {Milman}},\ and\ \bibinfo {author}
			{\bibfnamefont {S.}~\bibnamefont {Felicetti}},\ }\bibfield  {title} {\bibinfo
			{title} {Superradiant phase transition in the ultrastrong-coupling regime of
				the two-photon {Dicke} model},\ }\href
		{https://doi.org/10.1103/PhysRevA.95.053854} {\bibfield  {journal} {\bibinfo
				{journal} {Phys. Rev. A}\ }\textbf {\bibinfo {volume} {95}},\ \bibinfo
			{pages} {053854} (\bibinfo {year} {2017})}\BibitemShut {NoStop}%
		\bibitem [{\citenamefont {Cui}\ \emph {et~al.}(2020)\citenamefont {Cui},
			\citenamefont {Grémaud}, \citenamefont {Guo},\ and\ \citenamefont
			{Batrouni}}]{cui_nonlinear_2020}%
		\BibitemOpen
		\bibfield  {author} {\bibinfo {author} {\bibfnamefont {S.}~\bibnamefont
				{Cui}}, \bibinfo {author} {\bibfnamefont {B.}~\bibnamefont {Grémaud}},
			\bibinfo {author} {\bibfnamefont {W.}~\bibnamefont {Guo}},\ and\ \bibinfo
			{author} {\bibfnamefont {G.~G.}\ \bibnamefont {Batrouni}},\ }\bibfield
		{title} {\bibinfo {title} {Nonlinear two-photon {Rabi}-{Hubbard} model:
				superradiance, photon, and photon-pair {Bose}-{Einstein} condensates},\
		}\href {https://doi.org/10.1103/PhysRevA.102.033334} {\bibfield  {journal}
			{\bibinfo  {journal} {Phys. Rev. A}\ }\textbf {\bibinfo {volume} {102}},\
			\bibinfo {pages} {033334} (\bibinfo {year} {2020})}\BibitemShut {NoStop}%
		\bibitem [{\citenamefont {Felicetti}\ \emph
			{et~al.}(2018{\natexlab{a}})\citenamefont {Felicetti}, \citenamefont
			{Rossatto}, \citenamefont {Rico}, \citenamefont {Solano},\ and\ \citenamefont
			{Forn-Díaz}}]{felicetti_two-photon_2018}%
		\BibitemOpen
		\bibfield  {author} {\bibinfo {author} {\bibfnamefont {S.}~\bibnamefont
				{Felicetti}}, \bibinfo {author} {\bibfnamefont {D.~Z.}\ \bibnamefont
				{Rossatto}}, \bibinfo {author} {\bibfnamefont {E.}~\bibnamefont {Rico}},
			\bibinfo {author} {\bibfnamefont {E.}~\bibnamefont {Solano}},\ and\ \bibinfo
			{author} {\bibfnamefont {P.}~\bibnamefont {Forn-Díaz}},\ }\bibfield  {title}
		{\bibinfo {title} {Two-photon quantum {Rabi} model with superconducting
				circuits},\ }\href {https://doi.org/10.1103/PhysRevA.97.013851} {\bibfield
			{journal} {\bibinfo  {journal} {Phys. Rev. A}\ }\textbf {\bibinfo {volume}
				{97}},\ \bibinfo {pages} {013851} (\bibinfo {year}
			{2018}{\natexlab{a}})}\BibitemShut {NoStop}%
		\bibitem [{\citenamefont {Felicetti}\ \emph
			{et~al.}(2018{\natexlab{b}})\citenamefont {Felicetti}, \citenamefont
			{Hwang},\ and\ \citenamefont
			{Le~Boité}}]{felicetti_ultrastrong-coupling_2018}%
		\BibitemOpen
		\bibfield  {author} {\bibinfo {author} {\bibfnamefont {S.}~\bibnamefont
				{Felicetti}}, \bibinfo {author} {\bibfnamefont {M.-J.}\ \bibnamefont
				{Hwang}},\ and\ \bibinfo {author} {\bibfnamefont {A.}~\bibnamefont
				{Le~Boité}},\ }\bibfield  {title} {\bibinfo {title} {Ultrastrong-coupling
				regime of nondipolar light-matter interactions},\ }\href
		{https://doi.org/10.1103/PhysRevA.98.053859} {\bibfield  {journal} {\bibinfo
				{journal} {Phys. Rev. A}\ }\textbf {\bibinfo {volume} {98}},\ \bibinfo
			{pages} {053859} (\bibinfo {year} {2018}{\natexlab{b}})}\BibitemShut
		{NoStop}%
		\bibitem [{\citenamefont {Wang}\ \emph {et~al.}(2025)\citenamefont {Wang},
			\citenamefont {Mercurio}, \citenamefont {Ridolfo}, \citenamefont {Wang},
			\citenamefont {Chen}, \citenamefont {Wang}, \citenamefont {Liu},
			\citenamefont {Sun}, \citenamefont {Li}, \citenamefont {Nori}, \citenamefont
			{Savasta},\ and\ \citenamefont {You}}]{wang_strong_2025}%
		\BibitemOpen
		\bibfield  {author} {\bibinfo {author} {\bibfnamefont {S.-P.}\ \bibnamefont
				{Wang}}, \bibinfo {author} {\bibfnamefont {A.}~\bibnamefont {Mercurio}},
			\bibinfo {author} {\bibfnamefont {A.}~\bibnamefont {Ridolfo}}, \bibinfo
			{author} {\bibfnamefont {Y.}~\bibnamefont {Wang}}, \bibinfo {author}
			{\bibfnamefont {M.}~\bibnamefont {Chen}}, \bibinfo {author} {\bibfnamefont
				{W.}~\bibnamefont {Wang}}, \bibinfo {author} {\bibfnamefont {Y.}~\bibnamefont
				{Liu}}, \bibinfo {author} {\bibfnamefont {H.}~\bibnamefont {Sun}}, \bibinfo
			{author} {\bibfnamefont {T.}~\bibnamefont {Li}}, \bibinfo {author}
			{\bibfnamefont {F.}~\bibnamefont {Nori}}, \bibinfo {author} {\bibfnamefont
				{S.}~\bibnamefont {Savasta}},\ and\ \bibinfo {author} {\bibfnamefont {J.~Q.}\
				\bibnamefont {You}},\ }\bibfield  {title} {\bibinfo {title} {Strong coupling
				between a single-photon and a two-photon {Fock} state},\ }\href
		{https://doi.org/10.1038/s41467-025-63783-7} {\bibfield  {journal} {\bibinfo
				{journal} {Nat. Commun.}\ }\textbf {\bibinfo {volume} {16}},\ \bibinfo
			{pages} {8730} (\bibinfo {year} {2025})}\BibitemShut {NoStop}%
		\bibitem [{\citenamefont {Del~Valle}\ \emph {et~al.}(2010)\citenamefont
			{Del~Valle}, \citenamefont {Zippilli}, \citenamefont {Laussy}, \citenamefont
			{Gonzalez-Tudela}, \citenamefont {Morigi},\ and\ \citenamefont
			{Tejedor}}]{del_valle_two-photon_2010}%
		\BibitemOpen
		\bibfield  {author} {\bibinfo {author} {\bibfnamefont {E.}~\bibnamefont
				{Del~Valle}}, \bibinfo {author} {\bibfnamefont {S.}~\bibnamefont {Zippilli}},
			\bibinfo {author} {\bibfnamefont {F.~P.}\ \bibnamefont {Laussy}}, \bibinfo
			{author} {\bibfnamefont {A.}~\bibnamefont {Gonzalez-Tudela}}, \bibinfo
			{author} {\bibfnamefont {G.}~\bibnamefont {Morigi}},\ and\ \bibinfo {author}
			{\bibfnamefont {C.}~\bibnamefont {Tejedor}},\ }\bibfield  {title} {\bibinfo
			{title} {Two-photon lasing by a single quantum dot in a high-{Q}
				microcavity},\ }\href {https://doi.org/10.1103/PhysRevB.81.035302} {\bibfield
			{journal} {\bibinfo  {journal} {Phys. Rev. B}\ }\textbf {\bibinfo {volume}
				{81}},\ \bibinfo {pages} {035302} (\bibinfo {year} {2010})}\BibitemShut
		{NoStop}%
		\bibitem [{\citenamefont {Felicetti}\ \emph {et~al.}(2015)\citenamefont
			{Felicetti}, \citenamefont {Pedernales}, \citenamefont {Egusquiza},
			\citenamefont {Romero}, \citenamefont {Lamata}, \citenamefont {Braak},\ and\
			\citenamefont {Solano}}]{felicetti_spectral_2015}%
		\BibitemOpen
		\bibfield  {author} {\bibinfo {author} {\bibfnamefont {S.}~\bibnamefont
				{Felicetti}}, \bibinfo {author} {\bibfnamefont {J.~S.}\ \bibnamefont
				{Pedernales}}, \bibinfo {author} {\bibfnamefont {I.~L.}\ \bibnamefont
				{Egusquiza}}, \bibinfo {author} {\bibfnamefont {G.}~\bibnamefont {Romero}},
			\bibinfo {author} {\bibfnamefont {L.}~\bibnamefont {Lamata}}, \bibinfo
			{author} {\bibfnamefont {D.}~\bibnamefont {Braak}},\ and\ \bibinfo {author}
			{\bibfnamefont {E.}~\bibnamefont {Solano}},\ }\bibfield  {title} {\bibinfo
			{title} {Spectral collapse via two-phonon interactions in trapped ions},\
		}\href {https://doi.org/10.1103/PhysRevA.92.033817} {\bibfield  {journal}
			{\bibinfo  {journal} {Phys. Rev. A}\ }\textbf {\bibinfo {volume} {92}},\
			\bibinfo {pages} {033817} (\bibinfo {year} {2015})}\BibitemShut {NoStop}%
		\bibitem [{\citenamefont {Puebla}\ \emph {et~al.}(2019)\citenamefont {Puebla},
			\citenamefont {Casanova}, \citenamefont {Houhou}, \citenamefont {Solano},\
			and\ \citenamefont {Paternostro}}]{puebla_quantum_2019}%
		\BibitemOpen
		\bibfield  {author} {\bibinfo {author} {\bibfnamefont {R.}~\bibnamefont
				{Puebla}}, \bibinfo {author} {\bibfnamefont {J.}~\bibnamefont {Casanova}},
			\bibinfo {author} {\bibfnamefont {O.}~\bibnamefont {Houhou}}, \bibinfo
			{author} {\bibfnamefont {E.}~\bibnamefont {Solano}},\ and\ \bibinfo {author}
			{\bibfnamefont {M.}~\bibnamefont {Paternostro}},\ }\bibfield  {title}
		{\bibinfo {title} {Quantum simulation of multiphoton and nonlinear
				dissipative spin-boson models},\ }\href
		{https://doi.org/10.1103/PhysRevA.99.032303} {\bibfield  {journal} {\bibinfo
				{journal} {Phys. Rev. A}\ }\textbf {\bibinfo {volume} {99}},\ \bibinfo
			{pages} {032303} (\bibinfo {year} {2019})}\BibitemShut {NoStop}%
		\bibitem [{\citenamefont {Casanova}\ \emph {et~al.}(2018)\citenamefont
			{Casanova}, \citenamefont {Puebla}, \citenamefont {Moya-Cessa},\ and\
			\citenamefont {Plenio}}]{casanova_connecting_2018}%
		\BibitemOpen
		\bibfield  {author} {\bibinfo {author} {\bibfnamefont {J.}~\bibnamefont
				{Casanova}}, \bibinfo {author} {\bibfnamefont {R.}~\bibnamefont {Puebla}},
			\bibinfo {author} {\bibfnamefont {H.}~\bibnamefont {Moya-Cessa}},\ and\
			\bibinfo {author} {\bibfnamefont {M.~B.}\ \bibnamefont {Plenio}},\ }\bibfield
		{title} {\bibinfo {title} {Connecting nth order generalised quantum {Rabi}
				models: emergence of nonlinear spin-boson coupling via spin rotations},\
		}\href {https://doi.org/10.1038/s41534-018-0096-9} {\bibfield  {journal}
			{\bibinfo  {journal} {npj Quantum Inf.}\ }\textbf {\bibinfo {volume} {4}},\
			\bibinfo {pages} {47} (\bibinfo {year} {2018})}\BibitemShut {NoStop}%
		\bibitem [{\citenamefont {Villas-Boas}\ and\ \citenamefont
			{Rossatto}(2019)}]{villas-boas_multiphoton_2019}%
		\BibitemOpen
		\bibfield  {author} {\bibinfo {author} {\bibfnamefont {C.~J.}\ \bibnamefont
				{Villas-Boas}}\ and\ \bibinfo {author} {\bibfnamefont {D.~Z.}\ \bibnamefont
				{Rossatto}},\ }\bibfield  {title} {\bibinfo {title} {Multiphoton
				{Jaynes}-{Cummings} model: arbitrary rotations in {Fock} space and quantum
				filters},\ }\href {https://doi.org/10.1103/PhysRevLett.122.123604} {\bibfield
			{journal} {\bibinfo  {journal} {Phys. Rev. Lett.}\ }\textbf {\bibinfo
				{volume} {122}},\ \bibinfo {pages} {123604} (\bibinfo {year}
			{2019})}\BibitemShut {NoStop}%
		\bibitem [{\citenamefont {Piccione}\ \emph {et~al.}(2022)\citenamefont
			{Piccione}, \citenamefont {Felicetti},\ and\ \citenamefont
			{Bellomo}}]{piccione_two-photon-interaction_2022}%
		\BibitemOpen
		\bibfield  {author} {\bibinfo {author} {\bibfnamefont {N.}~\bibnamefont
				{Piccione}}, \bibinfo {author} {\bibfnamefont {S.}~\bibnamefont
				{Felicetti}},\ and\ \bibinfo {author} {\bibfnamefont {B.}~\bibnamefont
				{Bellomo}},\ }\bibfield  {title} {\bibinfo {title} {Two-photon-interaction
				effects in the bad-cavity limit},\ }\href
		{https://doi.org/10.1103/PhysRevA.105.L011702} {\bibfield  {journal}
			{\bibinfo  {journal} {Phys. Rev. A}\ }\textbf {\bibinfo {volume} {105}},\
			\bibinfo {pages} {L011702} (\bibinfo {year} {2022})}\BibitemShut {NoStop}%
		\bibitem [{\citenamefont {Chen}\ \emph {et~al.}(2012)\citenamefont {Chen},
			\citenamefont {Wang}, \citenamefont {He}, \citenamefont {Liu},\ and\
			\citenamefont {Wang}}]{chen_exact_2012}%
		\BibitemOpen
		\bibfield  {author} {\bibinfo {author} {\bibfnamefont {Q.-H.}\ \bibnamefont
				{Chen}}, \bibinfo {author} {\bibfnamefont {C.}~\bibnamefont {Wang}}, \bibinfo
			{author} {\bibfnamefont {S.}~\bibnamefont {He}}, \bibinfo {author}
			{\bibfnamefont {T.}~\bibnamefont {Liu}},\ and\ \bibinfo {author}
			{\bibfnamefont {K.-L.}\ \bibnamefont {Wang}},\ }\bibfield  {title} {\bibinfo
			{title} {Exact solvability of the quantum {Rabi} model using {Bogoliubov}
				operators},\ }\href {https://doi.org/10.1103/PhysRevA.86.023822} {\bibfield
			{journal} {\bibinfo  {journal} {Phys. Rev. A}\ }\textbf {\bibinfo {volume}
				{86}},\ \bibinfo {pages} {023822} (\bibinfo {year} {2012})}\BibitemShut
		{NoStop}%
		\bibitem [{\citenamefont {Rico}\ \emph {et~al.}(2020)\citenamefont {Rico},
			\citenamefont {Maldonado-Villamizar},\ and\ \citenamefont
			{Rodriguez-Lara}}]{rico_spectral_2020}%
		\BibitemOpen
		\bibfield  {author} {\bibinfo {author} {\bibfnamefont {R.~J.~A.}\
				\bibnamefont {Rico}}, \bibinfo {author} {\bibfnamefont {F.~H.}\ \bibnamefont
				{Maldonado-Villamizar}},\ and\ \bibinfo {author} {\bibfnamefont {B.~M.}\
				\bibnamefont {Rodriguez-Lara}},\ }\bibfield  {title} {\bibinfo {title}
			{Spectral collapse in the two-photon quantum {Rabi} model},\ }\href
		{https://doi.org/10.1103/PhysRevA.101.063825} {\bibfield  {journal} {\bibinfo
				{journal} {Phys. Rev. A}\ }\textbf {\bibinfo {volume} {101}},\ \bibinfo
			{pages} {063825} (\bibinfo {year} {2020})}\BibitemShut {NoStop}%
		\bibitem [{\citenamefont {Braak}(2023)}]{braak_spectral_2023}%
		\BibitemOpen
		\bibfield  {author} {\bibinfo {author} {\bibfnamefont {D.}~\bibnamefont
				{Braak}},\ }\bibfield  {title} {\bibinfo {title} {Spectral determinant of the
				two‐photon quantum {Rabi} model},\ }\href
		{https://doi.org/10.1002/andp.202200519} {\bibfield  {journal} {\bibinfo
				{journal} {Annalen der Physik}\ }\textbf {\bibinfo {volume} {535}},\ \bibinfo
			{pages} {2200519} (\bibinfo {year} {2023})}\BibitemShut {NoStop}%
		\bibitem [{\citenamefont {Ying}(2021)}]{ying_symmetry-breaking_2021}%
		\BibitemOpen
		\bibfield  {author} {\bibinfo {author} {\bibfnamefont {Z.-J.}\ \bibnamefont
				{Ying}},\ }\bibfield  {title} {\bibinfo {title} {Symmetry-breaking patterns,
				tricriticalities, and quadruple points in the quantum {Rabi} model with bias
				and nonlinear interaction},\ }\href
		{https://doi.org/10.1103/PhysRevA.103.063701} {\bibfield  {journal} {\bibinfo
				{journal} {Phys. Rev. A}\ }\textbf {\bibinfo {volume} {103}},\ \bibinfo
			{pages} {063701} (\bibinfo {year} {2021})}\BibitemShut {NoStop}%
		\bibitem [{\citenamefont {Hammani}\ \emph {et~al.}(2024)\citenamefont
			{Hammani}, \citenamefont {Sakhi},\ and\ \citenamefont
			{Bennai}}]{hammani_two-photon_2024}%
		\BibitemOpen
		\bibfield  {author} {\bibinfo {author} {\bibfnamefont {M.}~\bibnamefont
				{Hammani}}, \bibinfo {author} {\bibfnamefont {Z.}~\bibnamefont {Sakhi}},\
			and\ \bibinfo {author} {\bibfnamefont {M.}~\bibnamefont {Bennai}},\
		}\bibfield  {title} {\bibinfo {title} {On the two-photon quantum {Rabi} model
				at the critical coupling strength},\ }\href
		{https://doi.org/10.1007/s11082-023-05522-0} {\bibfield  {journal} {\bibinfo
				{journal} {Opt. Quant. Electron}\ }\textbf {\bibinfo {volume} {56}},\
			\bibinfo {pages} {102} (\bibinfo {year} {2024})}\BibitemShut {NoStop}%
		\bibitem [{\citenamefont {Ying}\ \emph {et~al.}(2025)\citenamefont {Ying},
			\citenamefont {Han}, \citenamefont {Li}, \citenamefont {Felicetti},\ and\
			\citenamefont {Braak}}]{ying_critical_2025}%
		\BibitemOpen
		\bibfield  {author} {\bibinfo {author} {\bibfnamefont {Z.}~\bibnamefont
				{Ying}}, \bibinfo {author} {\bibfnamefont {H.}~\bibnamefont {Han}}, \bibinfo
			{author} {\bibfnamefont {B.}~\bibnamefont {Li}}, \bibinfo {author}
			{\bibfnamefont {S.}~\bibnamefont {Felicetti}},\ and\ \bibinfo {author}
			{\bibfnamefont {D.}~\bibnamefont {Braak}},\ }\bibfield  {title} {\bibinfo
			{title} {Critical quantum metrology in a stabilized two‐photon {Rabi}
				model},\ }\href {https://doi.org/10.1002/qute.202500263} {\bibfield
			{journal} {\bibinfo  {journal} {Adv. Quantum Tech.}\ }\textbf {\bibinfo
				{volume} {8}},\ \bibinfo {pages} {e00263} (\bibinfo {year}
			{2025})}\BibitemShut {NoStop}%
		\bibitem [{\citenamefont {Li}\ \emph {et~al.}(2025)\citenamefont {Li},
			\citenamefont {Braak},\ and\ \citenamefont {Chen}}]{li_critical_2025}%
		\BibitemOpen
		\bibfield  {author} {\bibinfo {author} {\bibfnamefont {J.}~\bibnamefont
				{Li}}, \bibinfo {author} {\bibfnamefont {D.}~\bibnamefont {Braak}},\ and\
			\bibinfo {author} {\bibfnamefont {Q.-H.}\ \bibnamefont {Chen}},\ }\bibfield
		{title} {\bibinfo {title} {Critical spectrum of the anisotropic two-photon
				quantum {Rabi} model},\ }\href {https://doi.org/10.1103/PhysRevA.111.043706}
		{\bibfield  {journal} {\bibinfo  {journal} {Phys. Rev. A}\ }\textbf {\bibinfo
				{volume} {111}},\ \bibinfo {pages} {043706} (\bibinfo {year}
			{2025})}\BibitemShut {NoStop}%
		\bibitem [{\citenamefont {Zulli}\ \emph {et~al.}(2025)\citenamefont {Zulli},
			\citenamefont {Mulkerin}, \citenamefont {Parish},\ and\ \citenamefont
			{Levinsen}}]{zulli_universal_2025}%
		\BibitemOpen
		\bibfield  {author} {\bibinfo {author} {\bibfnamefont {A.~N.}\ \bibnamefont
				{Zulli}}, \bibinfo {author} {\bibfnamefont {B.~C.}\ \bibnamefont {Mulkerin}},
			\bibinfo {author} {\bibfnamefont {M.~M.}\ \bibnamefont {Parish}},\ and\
			\bibinfo {author} {\bibfnamefont {J.}~\bibnamefont {Levinsen}},\ }\bibfield
		{title} {\bibinfo {title} {Universal {Efimov} scaling in the {Rabi}-coupled
				few-body spectrum},\ }\href {https://doi.org/10.1103/yr6z-mtyq} {\bibfield
			{journal} {\bibinfo  {journal} {Phys. Rev. Lett.}\ }\textbf {\bibinfo
				{volume} {135}},\ \bibinfo {pages} {053401} (\bibinfo {year}
			{2025})}\BibitemShut {NoStop}%
		\bibitem [{\citenamefont {Belitz}\ and\ \citenamefont
			{Vojta}(2005)}]{belitzHowGenericScale2005}%
		\BibitemOpen
		\bibfield  {author} {\bibinfo {author} {\bibfnamefont {D.}~\bibnamefont
				{Belitz}}\ and\ \bibinfo {author} {\bibfnamefont {T.}~\bibnamefont {Vojta}},\
		}\bibfield  {title} {\bibinfo {title} {How generic scale invariance
				influences quantum and classical phase transitions},\ }\href
		{https://doi.org/10.1103/RevModPhys.77.579} {\bibfield  {journal} {\bibinfo
				{journal} {Rev. Mod. Phys.}\ }\textbf {\bibinfo {volume} {77}},\ \bibinfo
			{pages} {579} (\bibinfo {year} {2005})}\BibitemShut {NoStop}%
		\bibitem [{\citenamefont {Sachdev}(2015)}]{sachdev_quantum_2015}%
		\BibitemOpen
		\bibfield  {author} {\bibinfo {author} {\bibfnamefont {S.}~\bibnamefont
				{Sachdev}},\ }\href@noop {} {\emph {\bibinfo {title} {Quantum phase
					transitions}}},\ \bibinfo {edition} {second edition, 5th printing}\ ed.\
		(\bibinfo  {publisher} {Cambridge University Press},\ \bibinfo {address}
		{Cambridge},\ \bibinfo {year} {2015})\BibitemShut {NoStop}%
		\bibitem [{\citenamefont {Ferri}\ \emph {et~al.}(2021)\citenamefont {Ferri},
			\citenamefont {{Rosa-Medina}}, \citenamefont {Finger}, \citenamefont {Dogra},
			\citenamefont {Soriente}, \citenamefont {Zilberberg}, \citenamefont
			{Donner},\ and\ \citenamefont
			{Esslinger}}]{ferriEmergingDissipativePhases2021}%
		\BibitemOpen
		\bibfield  {author} {\bibinfo {author} {\bibfnamefont {F.}~\bibnamefont
				{Ferri}}, \bibinfo {author} {\bibfnamefont {R.}~\bibnamefont
				{{Rosa-Medina}}}, \bibinfo {author} {\bibfnamefont {F.}~\bibnamefont
				{Finger}}, \bibinfo {author} {\bibfnamefont {N.}~\bibnamefont {Dogra}},
			\bibinfo {author} {\bibfnamefont {M.}~\bibnamefont {Soriente}}, \bibinfo
			{author} {\bibfnamefont {O.}~\bibnamefont {Zilberberg}}, \bibinfo {author}
			{\bibfnamefont {T.}~\bibnamefont {Donner}},\ and\ \bibinfo {author}
			{\bibfnamefont {T.}~\bibnamefont {Esslinger}},\ }\bibfield  {title} {\bibinfo
			{title} {Emerging dissipative phases in a superradiant quantum gas with
				tunable decay},\ }\href {https://doi.org/10.1103/PhysRevX.11.041046}
		{\bibfield  {journal} {\bibinfo  {journal} {Phys. Rev. X}\ }\textbf {\bibinfo
				{volume} {11}},\ \bibinfo {pages} {041046} (\bibinfo {year}
			{2021})}\BibitemShut {NoStop}%
		\bibitem [{\citenamefont {Stamp}\ \emph {et~al.}(2024)\citenamefont {Stamp},
			\citenamefont {Silevitch}, \citenamefont {Libersky}, \citenamefont
			{McKenzie}, \citenamefont {Geim},\ and\ \citenamefont
			{Rosenbaum}}]{stampGallerySoftModes2024}%
		\BibitemOpen
		\bibfield  {author} {\bibinfo {author} {\bibfnamefont {P.~C.~E.}\
				\bibnamefont {Stamp}}, \bibinfo {author} {\bibfnamefont {D.~M.}\ \bibnamefont
				{Silevitch}}, \bibinfo {author} {\bibfnamefont {M.}~\bibnamefont {Libersky}},
			\bibinfo {author} {\bibfnamefont {R.}~\bibnamefont {McKenzie}}, \bibinfo
			{author} {\bibfnamefont {A.~A.}\ \bibnamefont {Geim}},\ and\ \bibinfo
			{author} {\bibfnamefont {T.~F.}\ \bibnamefont {Rosenbaum}},\ }\bibfield
		{title} {\bibinfo {title} {Gallery of soft modes: Theory and experiment at a
				ferromagnetic quantum phase transition},\ }\href
		{https://doi.org/10.1103/PhysRevB.110.134420} {\bibfield  {journal} {\bibinfo
				{journal} {Phys. Rev. B}\ }\textbf {\bibinfo {volume} {110}},\ \bibinfo
			{pages} {134420} (\bibinfo {year} {2024})}\BibitemShut {NoStop}%
		\bibitem [{sup()}]{supplement}%
		\BibitemOpen
		\href@noop {} {}\bibinfo {note} {See Supplemental Material for further
			explanation and details of the calculation, which includes Refs.~[49,
			50].}\BibitemShut {Stop}%
		\bibitem [{\citenamefont {Xie}\ \emph {et~al.}(2019)\citenamefont {Xie},
			\citenamefont {Duan},\ and\ \citenamefont {Chen}}]{xie_quantum_2019}%
		\BibitemOpen
		\bibfield  {author} {\bibinfo {author} {\bibfnamefont {Y.-F.}\ \bibnamefont
				{Xie}}, \bibinfo {author} {\bibfnamefont {L.}~\bibnamefont {Duan}},\ and\
			\bibinfo {author} {\bibfnamefont {Q.-H.}\ \bibnamefont {Chen}},\ }\bibfield
		{title} {\bibinfo {title} {Quantum {Rabi}–{Stark} model: solutions and
				exotic energy spectra},\ }\href {https://doi.org/10.1088/1751-8121/ab1cf6}
		{\bibfield  {journal} {\bibinfo  {journal} {J. Phys. A: Math. Theor.}\
			}\textbf {\bibinfo {volume} {52}},\ \bibinfo {pages} {245304} (\bibinfo
			{year} {2019})}\BibitemShut {NoStop}%
		\bibitem [{\citenamefont {Chan}(2020)}]{chanBoundStatesTwophoton2020}%
		\BibitemOpen
		\bibfield  {author} {\bibinfo {author} {\bibfnamefont {C.~K.}\ \bibnamefont
				{Chan}},\ }\bibfield  {title} {\bibinfo {title} {Bound states of two-photon
				{Rabi} model at the collapse point},\ }\href
		{https://doi.org/10.1088/1751-8121/aba3e0} {\bibfield  {journal} {\bibinfo
				{journal} {J. Phys. A: Math. Theor.}\ }\textbf {\bibinfo {volume} {53}},\
			\bibinfo {pages} {385303} (\bibinfo {year} {2020})}\BibitemShut {NoStop}%
		\bibitem [{\citenamefont {Pezzé}\ and\ \citenamefont
			{Smerzi}(2009)}]{pezze_entanglement_2009}%
		\BibitemOpen
		\bibfield  {author} {\bibinfo {author} {\bibfnamefont {L.}~\bibnamefont
				{Pezzé}}\ and\ \bibinfo {author} {\bibfnamefont {A.}~\bibnamefont
				{Smerzi}},\ }\bibfield  {title} {\bibinfo {title} {Entanglement, nonlinear
				dynamics, and the {Heisenberg} limit},\ }\href
		{https://doi.org/10.1103/PhysRevLett.102.100401} {\bibfield  {journal}
			{\bibinfo  {journal} {Phys. Rev. Lett.}\ }\textbf {\bibinfo {volume} {102}},\
			\bibinfo {pages} {100401} (\bibinfo {year} {2009})}\BibitemShut {NoStop}%
		\bibitem [{\citenamefont {Hauke}\ \emph {et~al.}(2016)\citenamefont {Hauke},
			\citenamefont {Heyl}, \citenamefont {Tagliacozzo},\ and\ \citenamefont
			{Zoller}}]{hauke_measuring_2016}%
		\BibitemOpen
		\bibfield  {author} {\bibinfo {author} {\bibfnamefont {P.}~\bibnamefont
				{Hauke}}, \bibinfo {author} {\bibfnamefont {M.}~\bibnamefont {Heyl}},
			\bibinfo {author} {\bibfnamefont {L.}~\bibnamefont {Tagliacozzo}},\ and\
			\bibinfo {author} {\bibfnamefont {P.}~\bibnamefont {Zoller}},\ }\bibfield
		{title} {\bibinfo {title} {Measuring multipartite entanglement through
				dynamic susceptibilities},\ }\href {https://doi.org/10.1038/nphys3700}
		{\bibfield  {journal} {\bibinfo  {journal} {Nat. Phys.}\ }\textbf {\bibinfo
				{volume} {12}},\ \bibinfo {pages} {778} (\bibinfo {year} {2016})}\BibitemShut
		{NoStop}%
		\bibitem [{\citenamefont {Garbe}\ \emph {et~al.}(2020)\citenamefont {Garbe},
			\citenamefont {Bina}, \citenamefont {Keller}, \citenamefont {Paris},\ and\
			\citenamefont {Felicetti}}]{garbe_critical_2020}%
		\BibitemOpen
		\bibfield  {author} {\bibinfo {author} {\bibfnamefont {L.}~\bibnamefont
				{Garbe}}, \bibinfo {author} {\bibfnamefont {M.}~\bibnamefont {Bina}},
			\bibinfo {author} {\bibfnamefont {A.}~\bibnamefont {Keller}}, \bibinfo
			{author} {\bibfnamefont {M.~G.}\ \bibnamefont {Paris}},\ and\ \bibinfo
			{author} {\bibfnamefont {S.}~\bibnamefont {Felicetti}},\ }\bibfield  {title}
		{\bibinfo {title} {Critical quantum metrology with a finite-component quantum
				phase transition},\ }\href {https://doi.org/10.1103/PhysRevLett.124.120504}
		{\bibfield  {journal} {\bibinfo  {journal} {Phys. Rev. Lett.}\ }\textbf
			{\bibinfo {volume} {124}},\ \bibinfo {pages} {120504} (\bibinfo {year}
			{2020})}\BibitemShut {NoStop}%
		\bibitem [{\citenamefont {Chu}\ \emph {et~al.}(2021)\citenamefont {Chu},
			\citenamefont {Zhang}, \citenamefont {Yu},\ and\ \citenamefont
			{Cai}}]{chu_dynamic_2021}%
		\BibitemOpen
		\bibfield  {author} {\bibinfo {author} {\bibfnamefont {Y.}~\bibnamefont
				{Chu}}, \bibinfo {author} {\bibfnamefont {S.}~\bibnamefont {Zhang}}, \bibinfo
			{author} {\bibfnamefont {B.}~\bibnamefont {Yu}},\ and\ \bibinfo {author}
			{\bibfnamefont {J.}~\bibnamefont {Cai}},\ }\bibfield  {title} {\bibinfo
			{title} {Dynamic framework for criticality-enhanced quantum sensing},\ }\href
		{https://doi.org/10.1103/PhysRevLett.126.010502} {\bibfield  {journal}
			{\bibinfo  {journal} {Phys. Rev. Lett.}\ }\textbf {\bibinfo {volume} {126}},\
			\bibinfo {pages} {010502} (\bibinfo {year} {2021})}\BibitemShut {NoStop}%
		\bibitem [{\citenamefont {Kibble}(1976)}]{kibble_topology_1976}%
		\BibitemOpen
		\bibfield  {author} {\bibinfo {author} {\bibfnamefont {T.~W.~B.}\
				\bibnamefont {Kibble}},\ }\bibfield  {title} {\bibinfo {title} {Topology of
				cosmic domains and strings},\ }\href
		{https://doi.org/10.1088/0305-4470/9/8/029} {\bibfield  {journal} {\bibinfo
				{journal} {J. Phys. A: Math. Gen.}\ }\textbf {\bibinfo {volume} {9}},\
			\bibinfo {pages} {1387} (\bibinfo {year} {1976})}\BibitemShut {NoStop}%
		\bibitem [{\citenamefont {Zurek}(1985)}]{zurek_cosmological_1985}%
		\BibitemOpen
		\bibfield  {author} {\bibinfo {author} {\bibfnamefont {W.~H.}\ \bibnamefont
				{Zurek}},\ }\bibfield  {title} {\bibinfo {title} {Cosmological experiments in
				superfluid helium?},\ }\href {https://doi.org/10.1038/317505a0} {\bibfield
			{journal} {\bibinfo  {journal} {Nature}\ }\textbf {\bibinfo {volume} {317}},\
			\bibinfo {pages} {505} (\bibinfo {year} {1985})}\BibitemShut {NoStop}%
		\bibitem [{\citenamefont {Dziarmaga}(2010)}]{dziarmaga_dynamics_2010}%
		\BibitemOpen
		\bibfield  {author} {\bibinfo {author} {\bibfnamefont {J.}~\bibnamefont
				{Dziarmaga}},\ }\bibfield  {title} {\bibinfo {title} {Dynamics of a quantum
				phase transition and relaxation to a steady state},\ }\href
		{https://doi.org/10.1080/00018732.2010.514702} {\bibfield  {journal}
			{\bibinfo  {journal} {Adv. Phys.}\ }\textbf {\bibinfo {volume} {59}},\
			\bibinfo {pages} {1063} (\bibinfo {year} {2010})}\BibitemShut {NoStop}%
		\bibitem [{\citenamefont {Acevedo}\ \emph {et~al.}(2014)\citenamefont
			{Acevedo}, \citenamefont {Quiroga}, \citenamefont {Rodríguez},\ and\
			\citenamefont {Johnson}}]{acevedo_new_2014}%
		\BibitemOpen
		\bibfield  {author} {\bibinfo {author} {\bibfnamefont {O.}~\bibnamefont
				{Acevedo}}, \bibinfo {author} {\bibfnamefont {L.}~\bibnamefont {Quiroga}},
			\bibinfo {author} {\bibfnamefont {F.}~\bibnamefont {Rodríguez}},\ and\
			\bibinfo {author} {\bibfnamefont {N.}~\bibnamefont {Johnson}},\ }\bibfield
		{title} {\bibinfo {title} {New dynamical scaling universality for quantum
				networks across adiabatic quantum phase transitions},\ }\href
		{https://doi.org/10.1103/PhysRevLett.112.030403} {\bibfield  {journal}
			{\bibinfo  {journal} {Phys. Rev. Lett.}\ }\textbf {\bibinfo {volume} {112}},\
			\bibinfo {pages} {030403} (\bibinfo {year} {2014})}\BibitemShut {NoStop}%
		\bibitem [{\citenamefont {Defenu}\ \emph {et~al.}(2018)\citenamefont {Defenu},
			\citenamefont {Enss}, \citenamefont {Kastner},\ and\ \citenamefont
			{Morigi}}]{defenu_dynamical_2018}%
		\BibitemOpen
		\bibfield  {author} {\bibinfo {author} {\bibfnamefont {N.}~\bibnamefont
				{Defenu}}, \bibinfo {author} {\bibfnamefont {T.}~\bibnamefont {Enss}},
			\bibinfo {author} {\bibfnamefont {M.}~\bibnamefont {Kastner}},\ and\ \bibinfo
			{author} {\bibfnamefont {G.}~\bibnamefont {Morigi}},\ }\bibfield  {title}
		{\bibinfo {title} {Dynamical critical scaling of long-range interacting
				quantum magnets},\ }\href {https://doi.org/10.1103/PhysRevLett.121.240403}
		{\bibfield  {journal} {\bibinfo  {journal} {Phys. Rev. Lett.}\ }\textbf
			{\bibinfo {volume} {121}},\ \bibinfo {pages} {240403} (\bibinfo {year}
			{2018})}\BibitemShut {NoStop}%
		\bibitem [{\citenamefont {Lambert}\ \emph {et~al.}(2026)\citenamefont
			{Lambert}, \citenamefont {Giguere}, \citenamefont {Menczel}, \citenamefont
			{Li}, \citenamefont {Hopf}, \citenamefont {Suarez}, \citenamefont {Gali},
			\citenamefont {Lishman}, \citenamefont {Gadhvi}, \citenamefont {Agarwal},
			\citenamefont {Galicia}, \citenamefont {Shammah}, \citenamefont {Nation},
			\citenamefont {Johansson}, \citenamefont {Ahmed}, \citenamefont {Cross},
			\citenamefont {Pitchford},\ and\ \citenamefont {Nori}}]{qutip5}%
		\BibitemOpen
		\bibfield  {author} {\bibinfo {author} {\bibfnamefont {N.}~\bibnamefont
				{Lambert}}, \bibinfo {author} {\bibfnamefont {E.}~\bibnamefont {Giguere}},
			\bibinfo {author} {\bibfnamefont {P.}~\bibnamefont {Menczel}}, \bibinfo
			{author} {\bibfnamefont {B.}~\bibnamefont {Li}}, \bibinfo {author}
			{\bibfnamefont {P.}~\bibnamefont {Hopf}}, \bibinfo {author} {\bibfnamefont
				{G.}~\bibnamefont {Suarez}}, \bibinfo {author} {\bibfnamefont
				{M.}~\bibnamefont {Gali}}, \bibinfo {author} {\bibfnamefont {J.}~\bibnamefont
				{Lishman}}, \bibinfo {author} {\bibfnamefont {R.}~\bibnamefont {Gadhvi}},
			\bibinfo {author} {\bibfnamefont {R.}~\bibnamefont {Agarwal}}, \bibinfo
			{author} {\bibfnamefont {A.}~\bibnamefont {Galicia}}, \bibinfo {author}
			{\bibfnamefont {N.}~\bibnamefont {Shammah}}, \bibinfo {author} {\bibfnamefont
				{P.}~\bibnamefont {Nation}}, \bibinfo {author} {\bibfnamefont {J.~R.}\
				\bibnamefont {Johansson}}, \bibinfo {author} {\bibfnamefont {S.}~\bibnamefont
				{Ahmed}}, \bibinfo {author} {\bibfnamefont {S.}~\bibnamefont {Cross}},
			\bibinfo {author} {\bibfnamefont {A.}~\bibnamefont {Pitchford}},\ and\
			\bibinfo {author} {\bibfnamefont {F.}~\bibnamefont {Nori}},\ }\bibfield
		{title} {\bibinfo {title} {{QuTiP} 5: The quantum toolbox in {Python}},\
		}\href {https://doi.org/10.1016/j.physrep.2025.10.001} {\bibfield  {journal}
			{\bibinfo  {journal} {Phys. Rep.}\ }\textbf {\bibinfo {volume} {1153}},\
			\bibinfo {pages} {1} (\bibinfo {year} {2026})}\BibitemShut {NoStop}%
		\bibitem [{\citenamefont {Emary}\ and\ \citenamefont
			{Brandes}(2003)}]{emary_quantum_2003}%
		\BibitemOpen
		\bibfield  {author} {\bibinfo {author} {\bibfnamefont {C.}~\bibnamefont
				{Emary}}\ and\ \bibinfo {author} {\bibfnamefont {T.}~\bibnamefont
				{Brandes}},\ }\bibfield  {title} {\bibinfo {title} {Quantum chaos triggered
				by precursors of a quantum phase transition: the {Dicke} model},\ }\href
		{https://doi.org/10.1103/PhysRevLett.90.044101} {\bibfield  {journal}
			{\bibinfo  {journal} {Phys. Rev. Lett.}\ }\textbf {\bibinfo {volume} {90}},\
			\bibinfo {pages} {044101} (\bibinfo {year} {2003})}\BibitemShut {NoStop}%
		\bibitem [{\citenamefont {Chen}\ and\ \citenamefont
			{Zhang}(2018)}]{chen_finite-size_2018}%
		\BibitemOpen
		\bibfield  {author} {\bibinfo {author} {\bibfnamefont {X.-Y.}\ \bibnamefont
				{Chen}}\ and\ \bibinfo {author} {\bibfnamefont {Y.-Y.}\ \bibnamefont
				{Zhang}},\ }\bibfield  {title} {\bibinfo {title} {Finite-size scaling
				analysis in the two-photon {Dicke} model},\ }\href
		{https://doi.org/10.1103/PhysRevA.97.053821} {\bibfield  {journal} {\bibinfo
				{journal} {Phys. Rev. A}\ }\textbf {\bibinfo {volume} {97}},\ \bibinfo
			{pages} {053821} (\bibinfo {year} {2018})}\BibitemShut {NoStop}%
	\end{thebibliography}
	
	%
	
\end{document}


\title{Quantum Criticality from Spectral Collapse in the Two-Photon Rabi Model\\
		-Supplemental Material-
	}
	
	\author{Jiong Li$^{1}$}
	\author{Jun-Ling Wang$^{2}$}
	\author{Qing-Hu Chen$^{2,1,}$}
	\email{qhchen@zju.edu.cn}
	\author{Hai-Qing Lin$^{1,}$}
	\email{hqlin@zju.edu.cn}
	
	\affiliation{
		$^1$ Institute for Advanced Study in Physics and School of Physics, Zhejiang University, Hangzhou 310058, China \\
		$^2$ Zhejiang Key Laboratory of Micro-Nano Quantum Chips and Quantum Control, School of Physics,
		Zhejiang University, Hangzhou 310058, China 
	}
	\date{\today}
	\maketitle
	
	\section{Asymptotic Low-Energy Expansion and Spectral Hierarchy} \label{sec_AA}
	
	This section presents the asymptotic expansion underlying the low-energy spectral hierarchy discussed in the main text. We derive the diagonal adiabatic approximation (AA) spectrum and the scaling of the same-parity and different-parity gaps, and elucidate the origins of their exponents.
	
	The Hamiltonian of the anisotropic two-photon quantum Rabi model (tpQRM) is 
	\begin{equation}
		H = - \frac{\Delta}{2} \sigma_x+\omega a^\dagger a + g \frac{1+r}{2} \sigma_{z} \left( a^2 + a^{\dagger 2} \right)  + g \frac{1-r}{2} i \sigma_y \left( a^2 - a^{\dagger 2} \right).
		\label{H_0}
	\end{equation}
	In the qubit basis, with $\omega=1$, this Hamiltonian can be written in block form as
		\begin{equation}
			H=
			\begin{pmatrix}
				H_{11} & H_{12} \\
				H_{21} & H_{22}
			\end{pmatrix},
		\end{equation}
		with
		\begin{eqnarray}
			H_{11} &=& a^\dagger a
			+ g\frac{1+r}{2}\left(a^2+a^{\dagger 2}\right), \nonumber \\
			H_{22} &=& a^\dagger a
			- g\frac{1+r}{2}\left(a^2+a^{\dagger 2}\right), \nonumber \\
			H_{12} &=& -\frac{\Delta}{2}
			+ g\frac{1-r}{2}\left(a^2-a^{\dagger 2}\right), \nonumber \\
			H_{21} &=& -\frac{\Delta}{2}
			- g\frac{1-r}{2}\left(a^2-a^{\dagger 2}\right).
	\end{eqnarray}
	Because the Hamiltonian preserves the $\mathbb{Z}_4$ symmetry generated by $\Pi = -\sigma_x \exp \left(i \frac{\pi}{2} a^\dagger a\right)$, an eigenstate in the even-photon subspace can be written as
	\begin{equation}
		\ket{\Phi} =
		\begin{bmatrix}
			\sum_{n=0}^{\infty} c_n S(-\theta) \ket{2n} \\
			\mp \sum_{n=0}^{\infty} (-1)^n c_n S(\theta) \ket{2n}
		\end{bmatrix},
		\label{phi}
	\end{equation}
	where the $\mp$ sign corresponds to parity eigenvalues $\pm 1$ in the even-photon $q=1/4$ subspace and $S(\theta) = \exp \left[ \frac{\theta}{2} \left( a^{\dagger 2} - a^2 \right)\right]$ is the squeezing operator.
	
	Introducing the squeezing parameter
	\begin{equation}
		\theta = \frac{1}{4}\ln \frac{1 + g/g_c}{1 - g/g_c}, \quad g_c = \frac{1}{1+r},
	\end{equation}
	the transformed diagonal blocks reduce to effective harmonic-oscillator Hamiltonians, hereafter referred to as the harmonic blocks:
	\begin{equation}
		S(\theta) H_{11} S(-\theta) = S(-\theta) H_{22} S(\theta) = \left( a^\dagger a + \frac{1}{2} \right) \beta - \frac{1}{2},
	\end{equation}
	where $\beta = \sqrt{1-g^2/g_c^2}$ is the effective oscillator frequency, which vanishes at the collapse point. Throughout this work, we approach the collapse point from $g<g_c$, where $\beta$ remains real and the effective quadratic Hamiltonian remains confining.
	
\begin{figure}[tbp]
		\includegraphics[width=1.0\linewidth]{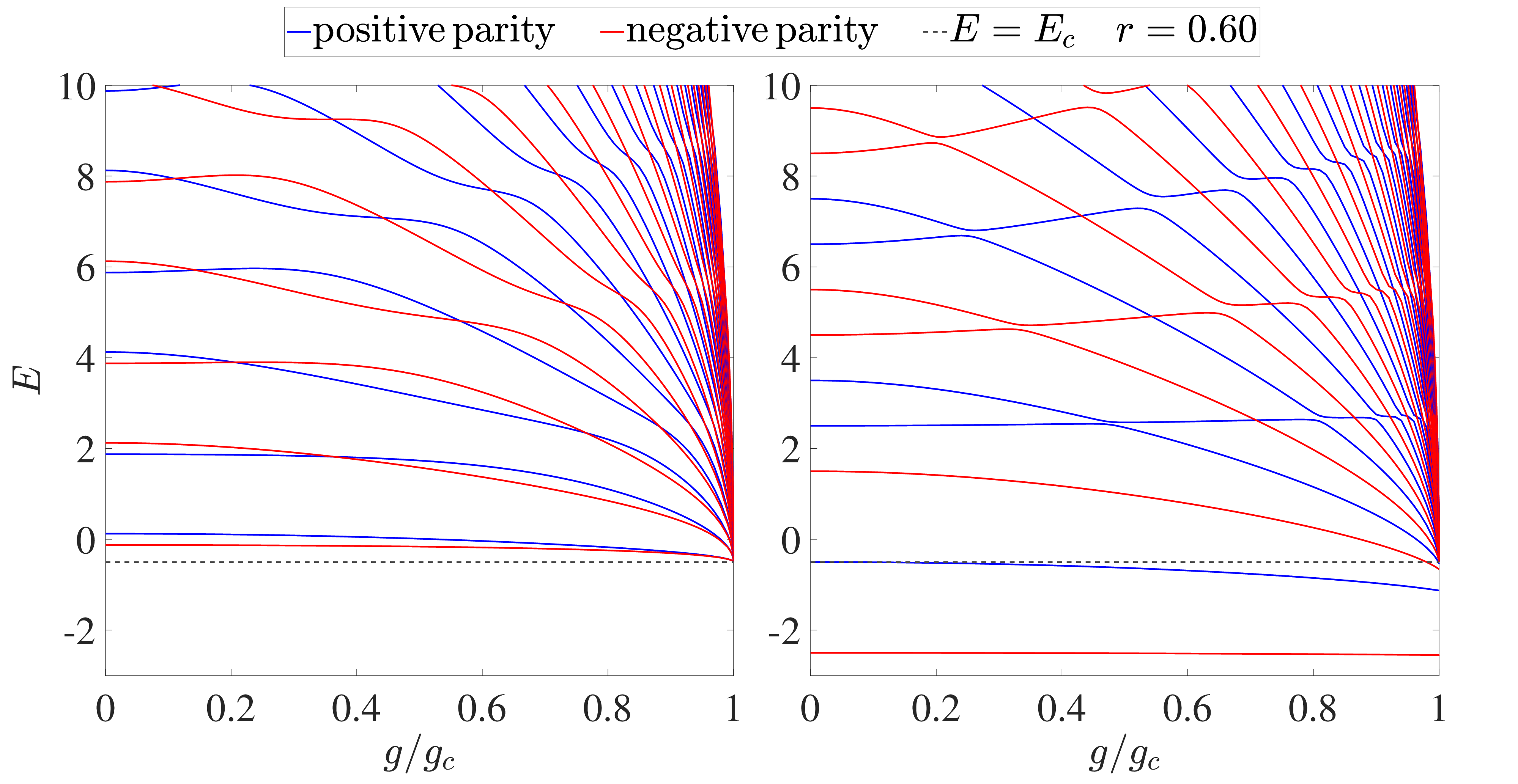}
		\caption{The parity-resolved spectrum for $r=0.60$ in the $q=1/4$ subspace. Blue and red curves denote positive and negative parity, respectively. The black dashed lines denote the continuum threshold $E_c$. Left: the critical condition $\Delta = \Delta_c = 0.25$. Right:  $\Delta = 5.0 \neq \Delta_c$. Under the critical condition, the discrete bound-state spectrum terminates at the continuum threshold. Away from the critical condition, discrete bound states remain below the continuum threshold.}
		\label{fig_full_spectra}
	\end{figure}

	Spectral collapse occurs at $g_c=(1+r)^{-1}$, at which $\beta$ vanishes and the discrete levels accumulate at the continuum threshold $E_c=-1/2$. To illustrate the evolution of bound states, Fig.~\ref{fig_full_spectra} shows the parity-resolved spectrum for $\Delta = \Delta_c =(1-r) / (1+r)$ and $\Delta \neq \Delta_c$. Consistent with Ref.~\cite{li_critical_2025}, at $\Delta = \Delta_c$, the discrete bound-state spectrum terminates at the continuum threshold. Away from the critical condition, by contrast, discrete bound states remain below the continuum threshold even at $g=g_c$. This distinction explains why spectral collapse constitutes a continuous QPT only under the critical condition.
	
	Projecting the upper component of the Schr\"odinger equation $H \ket{\Phi} = E \ket{\Phi}$ onto $\bra{2m} S(\theta)$ yields an infinite set of coupled equations,
	\begin{equation}
		0 = c_m \left[\left( 2m+\frac{1}{2} \right) \beta - \frac{1}{2} - E \right] \pm \sum_{n=0}^{\infty} c_n M_{mn}(\beta),
		\label{inf_matrix}
	\end{equation}
	which serve as the starting point for the AA. Here $M_{mn}(\beta)$ denotes a scalar matrix element of the off-diagonal block $H_{12}$ in the squeezed photonic basis, with the factor $(-1)^n$ in Eq.~\eqref{phi} absorbed into its definition:
		\begin{equation}
			M_{mn}(\beta) = -(-1)^n \bra{2m} S(\theta) H_{12} S(\theta)\ket{2n} = (-1)^n \bra{2m} S(\theta) \left[ \frac{\Delta}{2} - g \frac{1-r}{2} \left( a^2 - a^{\dagger 2} \right) \right] S(\theta) \ket{2n}.
	\end{equation}
	This matrix element can be expressed in terms of associated Legendre polynomials $P_l^k(x)$ as
	\begin{equation}
		M_{mn}(\beta) = (-1)^n \sqrt{\beta} \sqrt{\frac{(2n)!}{(2m)!}} \bigg[ \frac{\Delta}{2} P_{m+n}^{m-n} (\beta) - g \frac{1-r}{2} \left( P_{m+n-1}^{m-n+1} (\beta) - (2n+1)(2n+2) P_{m+n+1}^{m-n-1} (\beta) \right) \bigg],
	\end{equation}
	where we have used
	\begin{equation}
		\bra{2m} S(2\theta) \ket{2n} = \sqrt{\beta} \sqrt{\frac{ \left( 2n \right)! }{\left( 2m \right)!}} P_{m+n}^{m-n} (\beta), \quad \bra{2m} S(-2\theta) \ket{2n} = (-1)^{m-n} \sqrt{\beta} \sqrt{\frac{\left( 2n \right)!}{\left( 2m \right)!}} P_{m+n}^{m-n} (\beta).
	\end{equation}
	
	Within the AA, couplings between different squeezed-photon manifolds are neglected. The resulting eigenstates are
	\begin{equation}
		\ket{\psi_{n,\pm}^{(\rm AA)}} = \frac{1}{\sqrt{2}} \begin{bmatrix}
			S(-\theta) \ket{2n} \\
			\mp (-1)^n S(\theta) \ket{2n}
		\end{bmatrix},
	\end{equation}
	with eigenenergies
	\begin{equation}
		E_{n,\pm}^{\rm (AA)} = \left( 2n + \frac{1}{2} \right) \beta - \frac{1}{2} \pm M_{nn} (\beta) = \left( 2n + \frac{1}{2} \right) \beta - \frac{1}{2} \pm (-1)^n \sqrt{\beta} \left[ \frac{\Delta}{2} P_{2n}^{0} (\beta) - g \frac{1-r}{2} \left( P_{2n-1}^{1} (\beta) + P_{2n+1}^{1} (\beta) \right) \right].
	\end{equation}
	For $\beta \ll 1$, the associated Legendre polynomials admit the following asymptotic expansion
	\begin{equation}
		P_\ell^k (\beta) = \left[1 - (\ell+k+1) \frac{\ell-k}{2} \beta^2 + \mathcal{O} (\beta^4) \right] \frac{(\ell+k-1) !! (-1)^\frac{\ell-k}{2}}{\left( \ell-k \right)!!} (1-\beta^2)^\frac{k}{2},
	\end{equation}
	which gives
	\begin{equation}
		M_{mn}(\beta) = \frac{\sqrt{\beta}}{2} K_{mn}(\beta) \left[ \delta + \alpha_{mn} \beta^2 + \mathcal{O} (\beta^4) \right],
	\end{equation}
	where
	\begin{eqnarray}
		K_{mn}(\beta) &=& \sqrt{\frac{(2m-1)!! (2n-1)!!}{(2m)!! (2n)!!}} \left( 1-\beta^2 \right)^\frac{\vert m-n \vert}{2}, \nonumber \\
		\alpha_{mn} &=& (2m+1) \left( (5n+1) \Delta_c - n \Delta \right) - 2n \Delta_c,
		\nonumber \\
		\delta &=& \Delta - \Delta_c. 
	\end{eqnarray}	
	The overall prefactor $\sqrt{\beta}$ originates from the asymptotically vanishing overlap between the two oppositely squeezed oscillator branches, while $K_{mn}(\beta)$ introduces an additional suppression of off-diagonal couplings between different squeezed-photon manifolds.
	
	The asymptotic expansion yields the following AA spectrum near the collapse point [Eq.~(6) of the main text],
	\begin{equation}
		E_{n,\pm}^{(\rm AA)} \approx \left(2n + \frac{1}{2} \right) \beta - \frac{1}{2} \pm \frac{\sqrt{\beta}}{2} \frac{(2n-1)!!}{(2n)!!} \left( \delta + \alpha_{nn} \beta^{2} \right).
	\end{equation}
	Under the critical condition $\delta=0$, the same-parity and different-parity excitation gaps scale as [Eqs.~(7) and (8) of the main text], 
	\begin{eqnarray}
		\epsilon_{\rm sp} = \left| E_{n+1,\pm} - E_{n,\pm} \right| &\propto& \beta \propto |g - g_c|^{z\nu}, \quad z\nu = \frac{1}{2}, \nonumber\\
		\epsilon_{\rm dp} = \left| E_{n,+} - E_{n,-} \right| &\propto& \beta^{5/2} \propto |g - g_c|^{\mu}, \quad \mu = \frac{5}{4}.
	\end{eqnarray}
	These results demonstrate the asymptotic low-energy hierarchy discussed in the main text; the leading corrections beyond the diagonal AA are derived in the next section. The same-parity gap $\epsilon_{\rm sp}$ is set by the vanishing effective oscillator frequency, whereas the different-parity gap $\epsilon_{\rm dp}$ represents a symmetry-induced splitting. The exponent $\mu=5/4$ can be understood directly from the asymptotic expansion. The symmetry fixes the relation between the two spinor components in Eq.~\eqref{phi}, independent of the AA. Within the AA, the diagonal matrix element $M_{nn}$ quantifies the residual coupling between the two oppositely squeezed branches within the same squeezed-photon manifold, which produces the different-parity splitting. It has the asymptotic form
	\begin{equation}
		M_{nn} \propto \sqrt{\beta} \left( \delta + \alpha_{nn} \beta^2 + \mathcal{O}(\beta^4) \right).
	\end{equation}
	Away from the critical condition, the leading contribution is proportional to $\sqrt{\beta} \delta$. Under the critical condition, however, this term vanishes because $\delta=0$, so the leading nonzero contribution is the $\mathcal{O}(\beta^2)$ term in the small-$\beta$ expansion,
	\begin{equation}
		M_{nn} \propto \sqrt{\beta} \times \beta^2 = \beta^{5/2}.
	\end{equation}
	The symmetry-induced splitting therefore scales as
	\begin{equation}
		\epsilon_{\rm dp} = |E_{n,+} - E_{n,-}| \propto |M_{nn}| \propto \beta^{5/2} \propto |g-g_c|^{5/4}.
	\end{equation}
	Hence, the exponent $\mu=5/4$ characterizes the asymptotic suppression of the symmetry-induced splitting, rather than directly the softening of the effective oscillator mode.
	
	The hierarchy motivates the lowest same-parity excitation as the natural candidate for the soft mode, but does not by itself establish soft-mode control. That identification follows from the quantum Fisher information (QFI) response and Kibble-Zurek (KZ) dynamics analyzed in the main text.
	
	\section{Asymptotic Validity and Crossover Scale} \label{sec_validity}
	
	This section demonstrates the asymptotic validity of the AA. We show that the AA becomes a controlled asymptotic low-energy description under the critical condition $\Delta = \Delta_c$, whereas nonzero $\delta = \Delta-\Delta_c$ introduces a crossover scale $\delta^*(\beta) \propto \sqrt{\beta}$ beyond which off-diagonal couplings can no longer be neglected.
	
\begin{figure}[tbp]
		\includegraphics[width=1.0\linewidth]{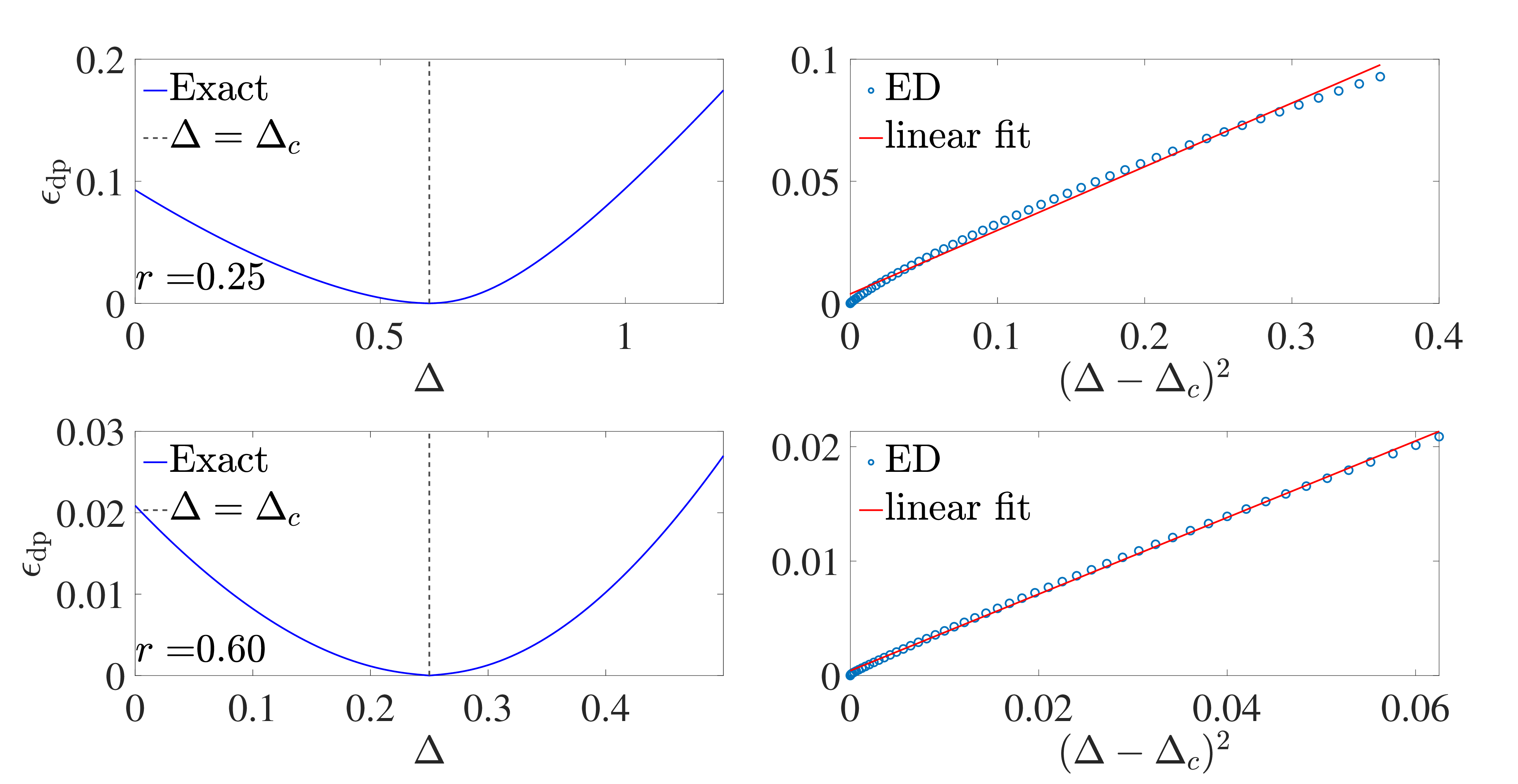}
		\caption{Left: The lowest gap in the full spectrum $\epsilon_{\rm dp}$ at $g=g_{c}$ as a function of $\Delta$, obtained by exact diagonalization for $r=0.25$ (upper panels) and $r=0.60$ (lower panels). The black dashed line marks the critical value $\Delta_c$. Right: The same data replotted as a function of $(\Delta - \Delta_c)^2$ for $\Delta < \Delta_c$ (open circles). The red line shows a linear fit, consistent with a quadratic opening near $\delta=0$.}
		\label{fig_energygap}
	\end{figure}

	To assess the validity of the AA, we treat the off-diagonal couplings perturbatively. The leading wavefunction corrections scale as
	\begin{equation}
		\ket{\psi_{n,\pm}^{(\rm off)}} = \sum_{m \neq n} \frac{M_{mn}(\beta) \ket{\psi_{m,\pm}^{(\rm AA)}}}{E_{n,\pm}^{(\rm AA)} - E_{m,\pm}^{(\rm AA)}}  \approx \sum_{m \neq n} (-1)^{m+n} \frac{ \sqrt{\beta} K_{mn}(\beta) \left( \delta + \alpha_{mn} \beta^2 \right)}{4(n-m) \beta} \ket{\psi_{m,\pm}^{(\rm AA)}} \propto
		\begin{cases}
			\mathcal{O}(\delta/\sqrt{\beta}), & \delta \neq 0, \\
			\mathcal{O}(\beta^{3/2}), & \delta = 0.
		\end{cases}
		\label{psi_off}
	\end{equation}
	Nonzero $\delta$ therefore induces strong off-diagonal couplings between different squeezed-photon manifolds as $\beta \to 0$, whereas the wavefunction correction vanishes asymptotically at $\delta=0$. The corresponding energy corrections scale as
	\begin{equation}
		E_{n,\pm}^{(\rm off)} = \sum_{m \neq n} \frac{\vert M_{mn} (\beta)\vert^2} { E_{n,\pm}^{(\rm AA)} - E_{m,\pm}^{(\rm AA)}} \approx \sum_{m \neq n} \frac{ \beta K_{mn}^2 (\beta) \left( \delta + \alpha_{mn} \beta^2 \right)^2}{8(n-m) \beta} \propto
		\begin{cases}
			\mathcal{O}(\delta^2), & \delta \neq 0, \\
			\mathcal{O}(\beta^4), & \delta = 0.
		\end{cases}
		\label{gap_off}
	\end{equation}
	Under the critical condition, the energy correction is of order $\mathcal{O}(\beta^4)$, and is therefore subleading relative to both $\epsilon_{\rm sp}$ and $\epsilon_{\rm dp}$. The critical exponents obtained within the AA are thus asymptotically exact. Consequently, the AA becomes a controlled asymptotic description for the low-energy eigenvalues and the leading asymptotic form of the corresponding eigenstates as $\beta \to 0$. Away from the critical condition, nonzero $\delta$ opens a finite lowest gap in the full spectrum of order $\mathcal{O}(\delta^2)$ at the collapse point, consistent with the exact diagonalization (ED) results in Fig.~\ref{fig_energygap}.
	
	We now examine the crossover induced by nonzero $\delta$. Although the AA becomes a controlled asymptotic description at $\delta=0$, nonzero $\delta$ introduces a finite gap at the collapse point and cuts off the critical scaling. From Eq.~\eqref{psi_off}, 
		\begin{equation}
			|\psi_{n,\pm}^{(\rm off)}| \propto \frac{|\delta|}{\sqrt{\beta}},
		\end{equation}
		so the AA remains self-consistent only when
		\begin{equation}
			|\delta| \ll \sqrt{\beta}.
		\end{equation}
		This defines the crossover scale 
		\begin{equation}
			\delta^*(\beta) \propto \sqrt{\beta}.
		\end{equation}
		For $|\delta| \gtrsim \delta^*(\beta)$, off-diagonal couplings become nonperturbative, and the asymptotic critical scaling crosses over to a noncritical regime in which the energy correction $E_{n,\pm}^{(\rm off)} \propto \delta^2$ cuts off the critical scaling.
	
\begin{figure}[tbp]
		\includegraphics[width=1.0\linewidth]{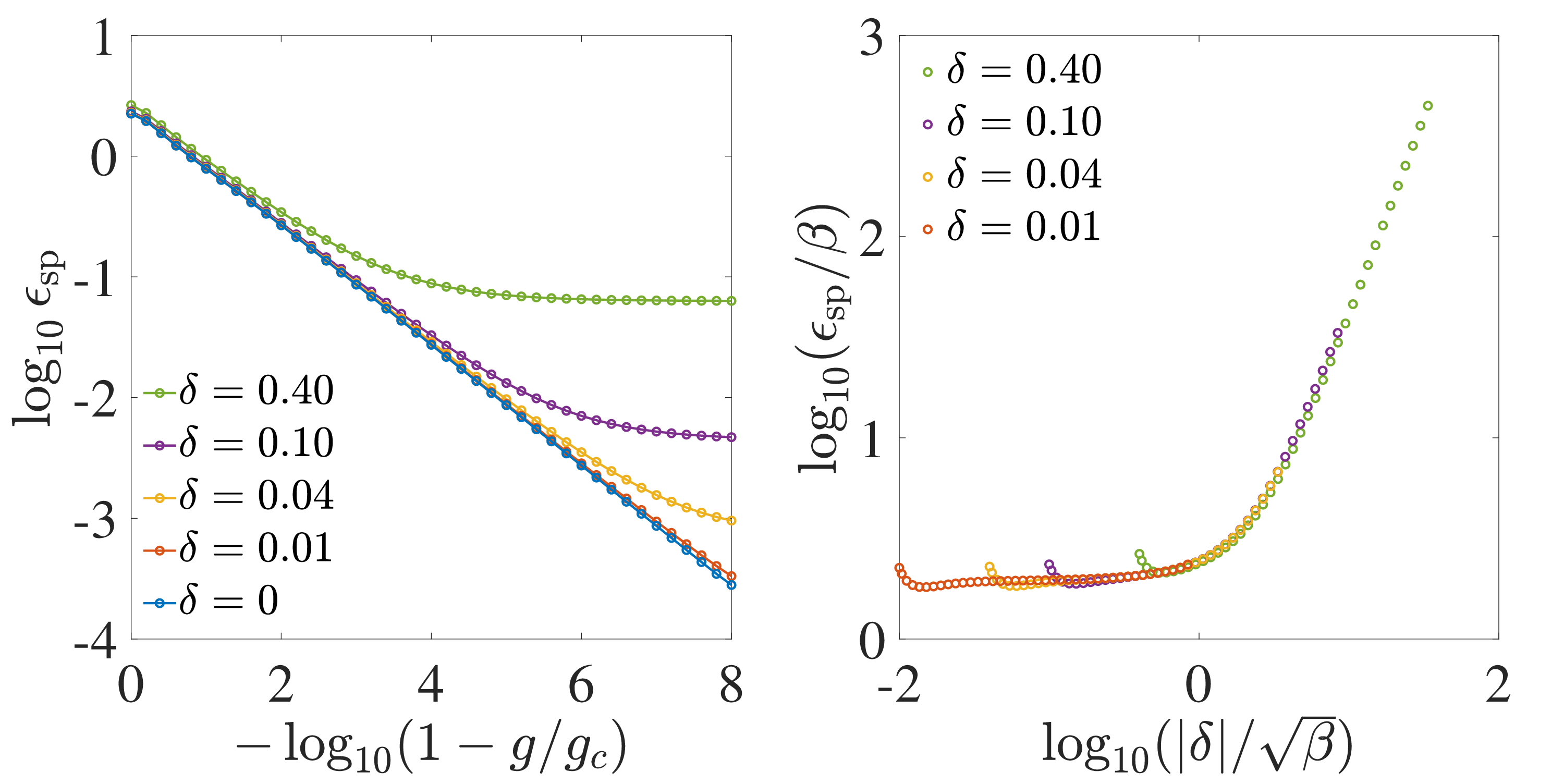}
		\caption{Verification of the predicted crossover by exact diagonalization at $r=0.60$. Left: The same-parity gap $\epsilon_{\rm sp}$ as a function of $-\log_{10}(1-g/g_c)$ for several values of $\delta$. Finite $\delta$ cuts off the gap closing near the collapse point. Right: Scaling collapse of $\epsilon_{\rm sp} / \beta$ replotted against $|\delta| / \sqrt{\beta}$. The crossover occurs near $|\delta| / \sqrt{\beta} \approx 1$, confirming the predicted crossover scale $\delta^*(\beta) \propto \sqrt{\beta}$.}
		\label{fig_crossover}
	\end{figure}

	Figure~\ref{fig_crossover} provides direct numerical confirmation of the predicted crossover. For $\delta \neq 0$, the same-parity gap deviates from its asymptotic critical scaling sufficiently close to the collapse point. When the data are replotted against the variable $|\delta| / \sqrt{\beta}$, they exhibit a clear scaling collapse, with the crossover occurring at $|\delta| / \sqrt{\beta} \approx 1$, providing direct numerical evidence for the crossover scale $\delta^*(\beta) \propto \sqrt{\beta}$. These results confirm that $\delta \neq 0$ cuts off the asymptotic critical scaling once $|\delta| \gtrsim \sqrt{\beta}$.
	
	Taken together, these results establish the asymptotic validity of the AA under the critical condition and quantify the crossover induced by nonzero $\delta$. Thus, the AA provides a controlled description only when $|\delta| \ll \sqrt{\beta}$; outside this regime, finite $\delta$ cuts off the critical scaling.
	
	\section{Spectral Collapse in the Isotropic tpQRM}
	
	This section complements the discussion of the isotropic limit presented in the main text. When $r=1$, the critical condition reduces to $\Delta_c=0$, so the different-parity gap vanishes identically. We show that as the collapse point is approached from the confining side $g<g_c$, the same-parity gap remains governed by the vanishing effective oscillator frequency and provides the low-energy scale associated with the soft mode identified in the main text.
	
	In the isotropic limit, the Hamiltonian reduces to
	\begin{equation}
		H_{\rm tp} = - \frac{\Delta}{2} \sigma_x+\omega a^\dagger a + g \sigma_{z} \left( a^2 + a^{\dagger 2} \right).
	\end{equation}
	Introducing the canonical position and momentum operators, $x = a + a^\dagger$ and $p = i(a^\dagger - a)$, the Hamiltonian becomes
	\begin{equation}
		H_{\rm tp} = \frac{1}{2} \begin{bmatrix}
			\left( \frac{1}{2} + g \right) x^2 + \left( \frac{1}{2} - g \right) p^2 - 1 & -\Delta \\
			-\Delta & \left( \frac{1}{2} - g \right) x^2 + \left( \frac{1}{2} + g \right) p^2 - 1
		\end{bmatrix},
		\label{Htp_x}
	\end{equation}
	where $p=-2i \partial_x$. This representation explicitly shows that the model contains a single bosonic mode. The two diagonal qubit sectors correspond to two distinct quadratic Hamiltonians of the same canonical variables $x$ and $p$, and they are coupled by the qubit tunneling term proportional to $\Delta$. The corresponding effective oscillator frequency is $\omega_{\rm eff} = \sqrt{1-4g^2}$, which vanishes at $g_c=1/2$.
	
	For $g<g_c$, the quadratic form in Eq.~\eqref{Htp_x} is confining and defines a stable harmonic-oscillator problem. As the collapse point is approached, $\omega_{\rm eff}$ continuously vanishes. By contrast, for $g>g_c$, the quadratic form becomes indefinite: in each diagonal block, one of the two quadrature coefficients becomes negative. Equivalently, $\omega_{\rm eff}$ becomes imaginary, indicating the loss of harmonic confinement and the disappearance of normalizable bound states, consistent with previous analyses of spectral collapse~\cite{felicetti_spectral_2015, braak_spectral_2023}. Consequently, the low-energy expansion and the AA developed in this work are applicable only when the collapse point is approached from the confining side, $g<g_c$. 
	
	In the Rabi-Stark model, the Hamiltonian simplifies substantially at the collapse point and can be solved analytically~\cite{xie_quantum_2019}. Similarly, for the tpQRM, the Hamiltonian at the collapse point $g = g_c$ reduces to
	\begin{equation}
		H_{\rm c} = \frac{1}{2} \begin{bmatrix}
			x^2 - 1 & -\Delta \\
			-\Delta &  p^2 - 1
		\end{bmatrix},
	\end{equation}
	where one canonical quadrature is completely unconfined. The resulting flattening of the effective potential along a canonical direction provides a direct phase-space illustration of spectral collapse as the loss of harmonic confinement.
	
	To analyze the resulting spectrum, we consider the Schr\"odinger equation for the spinor wavefunction $\varphi = [\varphi_1 \, \varphi_2]^{\rm T}$:
	\begin{equation}
		\frac{1}{2} \left( x^2 \varphi_1 - \Delta \varphi_2 \right) = \left( E+\frac{1}{2} \right) \varphi_1, \quad
		\frac{1}{2} \left( -\Delta \varphi_1 - 4 \partial_x^2 \varphi_2 \right) = \left( E+\frac{1}{2} \right) \varphi_2.
	\end{equation}	
	For $r=1$, the critical condition requires $\Delta_c = (1-r) / (1+r) = 0$. The two spinor components decouple, with one governed by the position operator and the other by the momentum operator. The corresponding generalized eigenstates are eigenfunctions of $x^2$ and $p^2$ with eigenvalues $x_0^2 = p_0^2 = 2E+1$, respectively. Since these eigenstates are non-normalizable, the spectrum is continuous above the continuum threshold $E_c=-1/2$, and no discrete bound states survive~\cite{braak_spectral_2023}.
	
	In this regime, the two qubit components are completely decoupled. If $[\varphi_1 \ \varphi_2]^{\rm T}$ is an eigenstate, then $[\varphi_1 \ -\varphi_2]^{\rm T}$ is also an eigenstate with opposite parity and the same energy. As a result, the different-parity gap $\epsilon_{\rm dp}$ vanishes identically, and the spectrum is doubly degenerate. This degeneracy, however, does not alter the soft-mode structure. As the collapse point is approached from $g<g_c$, the same-parity gap is governed by $\omega_{\rm eff}$, which vanishes continuously. Therefore, even in the isotropic limit, the vanishing same-parity gap on the confining side continues to set the characteristic low-energy scale associated with the soft mode.
	
	Away from the critical condition, for nonzero $\Delta$, eliminating $\varphi_1$ yields an effective equation for $\varphi_2$~\cite{chanBoundStatesTwophoton2020}:
	\begin{equation}
		- 2 \partial_x^2 \varphi_2 - \frac{\Delta^2}{4 \left(\frac{x^2}{2} - E -\frac{1}{2}\right)} \varphi_2 = \left( E+\frac{1}{2} \right) \varphi_2.
	\end{equation}
	For bound states with $E < -1/2$, introducing the rescaled variable $x \to \sqrt{2} \kappa x$ with $\kappa^2 = - \left( E + \frac{1}{2} \right) > 0$, transforms the equation into
	\begin{equation}
		- \partial_x^2 \varphi_2 + V \varphi_2 = - \kappa^4 \varphi_2, 
	\end{equation}
	where
	\begin{equation}
		V = - \frac{\Delta^2}{4 \left(x^2 + 1 \right)}.
	\end{equation}
	The effective potential exhibits an inverse-square asymptotic form $V(x) \propto -1/x^2$ as $x \to \infty$, which can support an infinite sequence of bound states accumulating toward the continuum threshold, 
		\begin{equation}
			|E_n-E_c| \propto \lambda^{-n}.
		\end{equation}
		Here $E_n$ denotes the energy of the $n$th bound state, $E_c=-1/2$ is the continuum threshold, and $\lambda>1$ is the discrete scaling factor~\cite{li_critical_2025}, analogous to Efimov scaling~\cite{zulli_universal_2025}. 
	
	Therefore, in the isotropic limit $r=1$ and under the critical condition $\Delta=0$, spectral collapse originates from the complete loss of confinement, producing a purely continuous spectrum above $E_c$. The vanishing of the different-parity gap originates from the exact symmetry-induced degeneracy discussed above, whereas the same-parity gap continues to be determined by the softening of the effective oscillator frequency. Consequently, the same-parity gap remains the relevant low-energy scale in the isotropic limit, consistent with the soft-mode identification established from the critical responses in the main text.
	
	\section{Equilibrium Critical Scaling within the Adiabatic Approximation}
	
	This section derives the equilibrium critical scaling relations discussed in the main text, including the quadrature fluctuations, mean photon occupation, atomic polarization, and QFI. At $\Delta=\Delta_c$ and $g \to g_c$, the AA provides a controlled approximation for deriving these scaling relations.
	
	Within the AA, the parity eigenstates at $\Delta = \Delta_c$ are
	\begin{equation}
		\ket{\psi_{n,\pm}^{(\rm AA)}}
		=
		\frac{1}{\sqrt2}
		\begin{bmatrix}
			S(-\theta)\ket{2n}\\
			\mp(-1)^nS(\theta)\ket{2n}
		\end{bmatrix},
	\end{equation}
	while the same-parity gap satisfies
	$\epsilon_{\rm sp} \propto \beta$.
	
	To derive the equilibrium scaling analytically, we first evaluate the quadrature fluctuations of the AA ground state. This derivation requires expressing the action of the bosonic operators on the ground state in the AA eigenbasis. The relevant operator identities are
	\begin{eqnarray}
		a^2 \ket{\psi_{0}} = \frac{1-\beta}{2\beta} \begin{bmatrix}
			S(-\theta) \ket{2} \\
			- S(\theta) \ket{2}
		\end{bmatrix} - \frac{\sqrt{1-\beta^2}}{2 \sqrt{2} \beta} \begin{bmatrix}
			S(-\theta) \ket{0} \\
			S(\theta) \ket{0}
		\end{bmatrix} = \frac{1-\beta}{\sqrt{2} \beta} \ket{\psi_{1,+}^{(\rm AA)}} - \frac{\sqrt{1-\beta^2}}{2 \beta} \ket{\psi_{0,+}^{(\rm AA)}}, \nonumber \\
		a^{\dagger 2} \ket{\psi_{0}} = \frac{1+\beta}{2\beta} \begin{bmatrix}
			S(-\theta) \ket{2} \\
			- S(\theta) \ket{2}
		\end{bmatrix} - \frac{\sqrt{1-\beta^2}}{2 \sqrt{2} \beta} \begin{bmatrix}
			S(-\theta) \ket{0} \\
			S(\theta) \ket{0}
		\end{bmatrix} = \frac{1+\beta}{\sqrt{2} \beta} \ket{\psi_{1,+}^{(\rm AA)}} - \frac{\sqrt{1-\beta^2}}{2 \beta} \ket{\psi_{0,+}^{(\rm AA)}}, \nonumber \\
		a^{\dagger} a \ket{\psi_{0}} = \frac{1-\beta}{2\sqrt{2} \beta} \begin{bmatrix}
			S(-\theta) \ket{0} \\
			- S(\theta) \ket{0}
		\end{bmatrix} - \frac{\sqrt{1-\beta^2}}{2\beta} \begin{bmatrix}
			S(-\theta) \ket{2} \\
			S(\theta) \ket{2}
		\end{bmatrix} = \frac{1-\beta}{2 \beta} \ket{\psi_{0,-}^{(\rm AA)}} - \frac{\sqrt{1-\beta^2}}{\sqrt{2} \beta} \ket{\psi_{1,-}^{(\rm AA)}}.
		\label{a_relation}
	\end{eqnarray}
	Here $\ket{\psi_0} \equiv \ket{\psi_{0,-}^{(\rm AA)}}$ denotes the AA ground state.
	
	The quadrature fluctuations are
	\begin{eqnarray}
		\Delta x &=& \sqrt{\braket{x^2} - \braket{x}^2} = \sqrt{\braket{a^2 + a^{\dagger 2} + 2a^{\dagger} a + 1}}, \nonumber \\
		\Delta p &=& \sqrt{\braket{p^2} - \braket{p}^2} = \sqrt{\braket{2a^{\dagger} a + 1 - a^2 - a^{\dagger 2}}},
	\end{eqnarray}
	where $\braket{x} = \braket{p} = 0$ because the ground state belongs to the even-photon subspace. Using Eq.~\eqref{a_relation} and the orthogonality of the AA eigenstates, we obtain
	\begin{eqnarray}
		\bra{\psi_{0}} a^2 \ket{\psi_{0}} = \frac{1-\beta}{\sqrt{2} \beta} \langle \psi_{0,-}^{(\rm AA)} \ket{\psi_{1,+}^{(\rm AA)}} - \frac{\sqrt{1-\beta^2}}{2 \beta} \langle \psi_{0,-}^{(\rm AA)} \ket{\psi_{0,+}^{(\rm AA)}} = 0, \nonumber \\
		\bra{\psi_{0}} a^{\dagger 2} \ket{\psi_{0}} = \frac{1+\beta}{\sqrt{2} \beta} \langle \psi_{0,-}^{(\rm AA)} \ket{\psi_{1,+}^{(\rm AA)}} - \frac{\sqrt{1-\beta^2}}{2 \beta} \langle \psi_{0,-}^{(\rm AA)} \ket{\psi_{0,+}^{(\rm AA)}} = 0, \nonumber \\
		\bra{\psi_{0}} a^{\dagger} a \ket{\psi_{0}} = \frac{1-\beta}{2 \beta} \langle \psi_{0,-}^{(\rm AA)} \ket{\psi_{0,-}^{(\rm AA)}} - \frac{\sqrt{1-\beta^2}}{\sqrt{2} \beta} \langle \psi_{0,-}^{(\rm AA)} \ket{\psi_{1,-}^{(\rm AA)}} = \frac{1-\beta}{2\beta}.
	\end{eqnarray}
	It follows that
	\begin{equation}
		\Delta x = \Delta p= \sqrt{ \frac{1-\beta}{\beta} + 1} = \beta^{-1/2} \propto \vert g - g_c \vert^{-\nu}, \quad \nu = 1/4,
	\end{equation}
	consistent with Eq.~(15) of the main text. 
	
	Other equilibrium observables exhibit corresponding scaling relations. The mean photon occupation scales as
	\begin{equation}
		\braket{a^\dagger a} = \frac{1-\beta}{2\beta} \propto |g-g_c|^{-1/2},
	\end{equation}
	reflecting the rapid growth of bosonic fluctuations as the effective oscillator softens into a strongly squeezed state, consistent with Eq.~(17) of the main text. The ground-state order parameter, namely the atomic polarization, scales as
	\begin{equation}
		\braket{\sigma_x} = \frac{1}{2} \left[ \bra{0} S(-2\theta) \ket{0} + \bra{0} S(2\theta) \ket{0}\right] = \sqrt{\beta} \propto |g-g_c|^{1/4},
	\end{equation}
	vanishing continuously as the qubit state approaches an equal-weight superposition, as expected for a continuous QPT and consistent with Eq.~(16) of the main text.
	
	We next derive the asymptotic scaling of the QFI and the associated parity selection rule. Its spectral representation is
	\begin{equation}
		F_Q = 4 \sum_{n \neq 0} \frac{\vert \bra{\Phi_n(g)} \partial_g H \ket{\Phi_0 (g)} \vert^2}{\left( E_n(g) - E_0 (g) \right)^2},
	\end{equation}
	where $H \ket{\Phi_n (g)} = E_n(g) \ket{\Phi_n (g)}$ and 
	\begin{equation}
		\partial_g H = \frac{1+r}{2} \sigma_{z} \left( a^2 + a^{\dagger 2} \right)  + \frac{1-r}{2} i \sigma_y \left( a^2 - a^{\dagger 2} \right).
	\end{equation} 
	Parity symmetry imposes the selection rule $[\Pi, \partial_g H]=0$, since
	\begin{equation}
		\Pi \partial_g H \Pi^{-1} = \frac{1+r}{2} (-\sigma_{z}) \left( (ia)^2 + (-ia^{\dagger})^2 \right)  + \frac{1-r}{2} (-i \sigma_y) \left( (ia)^2 - (-ia^{\dagger})^2 \right) = \partial_g H.
	\end{equation}
	Consequently, the spectral representation of the QFI contains only intermediate states with the same parity as the ground state. The leading asymptotic contribution therefore comes from the lowest same-parity excitation,
	\begin{equation} 
		F_Q \approx 4 \frac{\vert \bra{\psi_{1,-}^{(\rm AA)}} \partial_g H \ket{\psi_{0,-}^{(\rm AA)}} \vert^2}{\left( E_{1,-}^{\rm (AA)} - E_{0,-}^{\rm (AA)} \right)^2} = \frac{\vert \bra{\psi_{1,-}^{(\rm AA)}} (1+r) \sigma_{z} \left( a^2 + a^{\dagger 2} \right)  + (1-r) i \sigma_y \left( a^2 - a^{\dagger 2} \right) \ket{\psi_{0,-}^{(\rm AA)}} \vert^2}{(2\beta)^2}.
	\end{equation} 
	Using Eq.~\eqref{a_relation} and retaining the leading term as $\beta \to 0$,
	\begin{eqnarray}
		F_Q &\approx& \frac{\left \vert \bra{\psi_{1,-}^{(\rm AA)}} [(1+r) \sigma_{z} + (1-r) i \sigma_y] \frac{1-\beta}{\sqrt{2} \beta} \ket{\psi_{1,+}^{(\rm AA)}}  + \bra{\psi_{1,-}^{(\rm AA)}} [(1+r) \sigma_{z} - (1-r) i \sigma_y] \frac{1+\beta}{\sqrt{2} \beta} \ket{\psi_{1,+}^{(\rm AA)}} \right \vert^2}{(2\beta)^2} \nonumber \\
		&=& \frac{\left \vert \bra{\psi_{1,-}^{(\rm AA)}}(1+r) \sigma_{z} \frac{\sqrt{2}}{\beta} \ket{\psi_{1,+}^{(\rm AA)}}  - \bra{\psi_{1,-}^{(\rm AA)}} \sqrt{2}(1-r) i \sigma_y \ket{\psi_{1,+}^{(\rm AA)}} \right \vert^2}{(2\beta)^2} \approx \frac{(1+r)^2}{2\beta^4} \propto |g-g_c|^{-2},
	\end{eqnarray}
	which gives Eq.~(11) of the main text.
	
	For completeness, once $\epsilon_{\rm sp}$ is identified as the relevant excitation scale from the symmetry-resolved QFI response and KZ dynamics in the main text, the equilibrium scalings above can be expressed as:
	\begin{equation}
		\Delta x, \Delta p \propto \epsilon_{\rm sp}^{-1/2}, \quad
		\braket{a^\dagger a} \propto \epsilon_{\rm sp}^{-1}, \quad
		\braket{\sigma_x} \propto \epsilon_{\rm sp}^{1/2}, \quad
		F_Q \propto \epsilon_{\rm sp}^{-4}.
	\end{equation}

	\section{Kibble-Zurek Scaling of the Residual Energy}
	
	This section derives the KZ scaling of the residual energy presented in the main text. In particular, we show that the same-parity gap, $\epsilon_{\rm sp}$, is the gap entering the KZ freeze-out condition.
	
	Following the adiabatic perturbation theory of Refs.~\cite{dziarmaga_dynamics_2010, hwang_quantum_2015}, we derive the KZ scaling within the controlled low-energy theory established in the preceding sections. We consider a linear quench protocol $g(t) = g_f t / \tau_q = \dot{g} t$, where $g_f$ is the final coupling strength and $\dot{g} \ll 1$. Assuming that the system is initially prepared in the ground state at $g=0$, the wavefunction at $\Delta=\Delta_c$ can be expanded in the AA basis as
	\begin{equation}
		\ket{\Psi(t)} = \sum f_{n,\pm}(t) e^{-i \Theta_{n,\pm}(t)} \ket{\psi_{n,\pm}^{(\rm AA)}(t)},
	\end{equation}
	where $\Theta_{n,\pm}(t)=\int_0^t E_{n,\pm}^{(\rm AA)}(t') dt'$. Substituting this expansion into the Schr\"odinger equation gives
	\begin{equation}
		\dot{f}_{n,\pm} (t) = - \sum f_{m,\pm} (t) \dot{g} \bra{\psi_{n,\pm}^{\rm (AA)} (g(t))}  \partial_{g} \ket{\psi_{m,\pm}^{(\rm AA)} (g(t))} e^{i \left[\Theta_{n,\pm}(t) - \Theta_{m,\pm}(t)\right]}.
	\end{equation}
	Changing variables to $g = \dot{g} t$ gives
	\begin{equation}
		f_{n,\pm}(g) = - \sum \int_{0}^{g} f_{m,\pm} (g') \bra{\psi_{n,\pm}^{\rm (AA)} (g')} \partial_{g'} \ket{\psi_{m,\pm}^{(\rm AA)} (g')} e^{i \left[\Theta_{n,\pm}(g') - \Theta_{m,\pm}(g')\right]} \mathrm{d} g'.
	\end{equation}
	
	With the initial condition $f_{0,-}(0) = 1$, adiabatic perturbation theory~\cite{dziarmaga_dynamics_2010} yields
	\begin{equation}
		f_{n,\pm}(g) \approx - \int_{0}^{g} \bra{\psi_{n,\pm}^{\rm (AA)} (g')} \partial_{g'} \ket{\psi_{0,-}^{(\rm AA)} (g')} e^{i \left[\Theta_{n,\pm}(g') - \Theta_{0,-}(g')\right]} \mathrm{d} g'.
	\end{equation}
	For a sufficiently slow quench, the rapidly oscillating integral can be evaluated asymptotically by integration by parts,
	\begin{equation}
		\int_{x_1}^{x_2} f(x) e^{i A g(x)} \mathrm{d}x = \left. \frac{f(x)}{iA g'(x)}e^{i A g(x)} \right\vert_{x_1}^{x_2} + \mathcal{O}(A^{-2}), 
	\end{equation}
	which yields
	\begin{eqnarray}
		f_{n,\pm}(g) &\approx& \left. i\dot{g} \frac{\bra{\psi_{n,\pm}^{(\rm AA)} (g)} \partial_{g} \ket{\psi_{0,-}^{(\rm AA)} (g)}}{E_{n,\pm}^{(\rm AA)} (g) - E_{0,-}^{(\rm AA)} (g)} e^{i \left[\Theta_{n,\pm}(g) - \Theta_{0,-}(g)\right]} \right \vert_0^g + \mathcal{O} (\dot{g}^2) \nonumber \\
		&\approx& - i\dot{g} \frac{\sqrt{2}(1+r)}{8 \left( 1-g^2/g_c^2 \right)^{\frac{3}{2}}} e^{i \left[\Theta_{n,\pm}(g) - \Theta_{0,-}(g) \right]} \delta_{n,1} \delta_{\pm,-}.
	\end{eqnarray}
	Since $[H(g), \Pi] = 0$ throughout the quench, the instantaneous eigenstates and their $g$-derivatives remain within fixed parity, and only same-parity transitions occur. The dominant contribution comes from the lowest same-parity excitation, so the instantaneous gap $\eta$ that enters the KZ condition is identified with the same-parity gap. To leading order,
	\begin{equation}
		f_{1,-}(g) \approx - i\dot{g} \frac{\sqrt{2}(1+r)}{8 \left( 1-g^2/g_c^2 \right)^{\frac{3}{2}}} e^{i\left[ \Theta_{1,-}(g) - \Theta_{0,-}(g) \right]} .
	\end{equation}
	
	The residual energy is defined as the excess energy above the instantaneous ground-state energy at the end of the quench, 
	\begin{equation}
		E_r(g_f) \approx \epsilon_{\rm sp} \vert f_{1,-}(g_f) \vert^2
		\approx \tau_q^{-2} \frac{g_f^2 / g_c^2}{16 (1 - g_f^2 / g_c^2)^{5/2}}.		
		\label{Er_adiabatic}
	\end{equation}
	For a fixed final coupling sufficiently far from the critical point, this expression reduces to the adiabatic scaling $E_r\propto\tau_q^{-2}$.
	
	As the system approaches the critical point, the vanishing gap leads to a breakdown of adiabaticity. We denote by $g_K$ the freeze-out coupling at which the adiabatic evolution breaks down. Within the KZ framework~\cite{kibble_topology_1976, zurek_cosmological_1985}, this occurs when the intrinsic relaxation time $\tau_{\rm rel} \propto \eta^{-1}$ becomes comparable to the driving time scale $\tau_{\rm drive} \propto |\eta/\dot{\eta}|$, equivalent to $\eta^2 = |\dot{\eta}|$. Using the instantaneous gap $\eta = \epsilon_{\rm sp} \approx 2\beta$ and the linear quench $g(t)=g_f t/\tau_q$, this condition gives
	\begin{eqnarray}
		\eta^2(g_K) = 4 \left( 1- \frac{g_K^2}{g_c^2} \right) = 2 \frac{g_K g_f}{\tau_q} (1+r)^2 \left( 1- \frac{g_K^2}{g_c^2} \right)^{-1/2},\nonumber \\
		\left( 1-g_K/g_c \right)^{3/2} = \frac{g_K g_f}{2 \tau_q} (1+r)^2 \left( 1 + g_K/g_c \right)^{-3/2}.
	\end{eqnarray}
	Evaluating this condition in the critical limit $g_f \to g_c$ and $g_K \to g_c$ yields 
	\begin{eqnarray}
		1 - g_K/g_c &\approx& \left( \frac{2^{-3/2}}{2\tau_q} \right)^{2/3} = \left( \frac{2^{- 5/2}}{\tau_q} \right)^{2/3},
		\nonumber \\
		g_K / g_c &\approx& 1 - (4\sqrt{2} \tau_q)^{- 2/3}.
	\end{eqnarray}
	Substituting $g_K$ into Eq.~\eqref{Er_adiabatic} gives
	\begin{equation}
		E_r \approx \tau_q^{-2} \frac{g_K^2 / g_c^2}{16 (1 - g_K^2 / g_c^2)^{5/2}} 
		\approx \tau_q^{-2} \frac{(4\sqrt{2} \tau_q)^{5/3}}{16} 
		\propto \tau_q^{-1/3},
	\end{equation}
	which gives Eq.~(13) of the main text. For $z\nu = 1/2$, the result can be written equivalently as
	\begin{equation}
		1-g_K/g_c \propto \tau_q^{-1/(z\nu + 1)}, \quad E_r \propto \tau_q^{-z\nu/(z\nu + 1)},
	\end{equation}
	in agreement with the universal KZ scaling.
	
	These results demonstrate that the KZ scaling is governed by $\epsilon_{\rm sp}$. In particular, the same exponent $z\nu$ determines both the equilibrium scaling,
	\begin{equation}
		\epsilon_{\rm sp} \propto |g-g_c|^{z\nu},
	\end{equation}
	and the nonequilibrium KZ scaling,
	\begin{equation}
		E_r \propto \tau_q^{-\frac{z\nu}{1+z\nu}}.
	\end{equation}
	
	The KZ scaling derived above applies only within an intermediate quench-time window. The lower boundary follows from the validity of the low-energy description. Since the freeze-out gap scales as
	\begin{equation}
		\eta(g_K) \propto \tau_q^{-z\nu/ (z\nu + 1)},
	\end{equation}
	it must remain well below the characteristic gaps of noncritical higher-energy excitations. In the dimensionless units used here, this condition gives the parametric lower bound
	\begin{equation}
		\tau_q \gg 1.
	\end{equation}
	For faster quenches, higher-energy excitations contribute appreciably and obscure the KZ scaling.
	
	A second crossover arises for quenches terminating at a finite coupling $g_f<g_c$. To reach the freeze-out regime before the quench ends, the freeze-out point must satisfy
	\begin{equation}
		g_K < g_f.
	\end{equation}
	Since the same-parity gap decreases monotonically as $g$ approaches $g_c$, this condition is equivalent to
	\begin{equation}
		\eta(g_K) \gtrsim \eta(g_f).
	\end{equation}
	Using
	\begin{equation}
		1-g_K/g_c \propto \tau_q^{-1/(z\nu + 1)}
	\end{equation}
	and
	\begin{equation}
		\eta(g_f) \propto \left(1-g_f/g_c \right)^{z\nu},
	\end{equation}
	$\eta(g_K) \gtrsim \eta(g_f)$ yields
	\begin{equation}
		\tau_q \ll \left( 1-g_f/g_c \right)^{-(1+z\nu)}.
	\end{equation}
	For $z\nu=1/2$, the upper boundary becomes
	\begin{equation}
		\tau_q \ll \left(1-g_f/g_c \right)^{-3/2}.
	\end{equation}
	Combining these two bounds, the KZ scaling is expected within the approximate window
	\begin{equation}
		1 \ll \tau_q \ll \left(1-g_f/g_c \right)^{-3/2}.
	\end{equation}
	Below this window, noncritical higher-energy excitations obscure the KZ scaling. Above it, the quench terminates before the freeze-out regime is reached, and the residual energy crosses over to the adiabatic scaling $E_r \propto \tau_q^{-2}$. The upper boundary diverges as $g_f \to g_c$, so the observable KZ regime broadens as the final coupling approaches the critical point.
	
	
	%